\documentclass[aps,prd,preprintnumbers,showpacs,showkeys,nofootinbib,
superscriptaddress,fleqn,floatfix,tightenlines,10pt]{revtex4-2}
\usepackage{amsmath,amsfonts,amssymb,amscd,amsxtra,amsthm}
\usepackage{graphicx}  
\usepackage{epstopdf}
\usepackage{dcolumn}  
\usepackage{bm}          
\usepackage{hyperref}
\usepackage{slashed}
\usepackage{cancel}
\usepackage{float} 
\usepackage{mathtools}
\usepackage{amsbsy}
\usepackage{amstext}

\usepackage[utf8]{inputenc} 
\usepackage{booktabs}
\usepackage[normalem]{ulem} 
\usepackage[dvipsnames]{xcolor} 
\usepackage{tabularx}
\usepackage{enumitem}  
\usepackage{array} 
\usepackage{slashed}
\usepackage{tikz}
\usepackage{float}
\usepackage{multirow}
\usepackage{subcaption}
\usepackage{threeparttable}
\renewcommand\sout{\bgroup \color{red} \ULdepth=-.5ex \ULset}

\makeatletter

\begin{document}  
 \preprint{INHA-NTG-09/2022}
\title{Quark spin content of SU(3) light and singly heavy baryons}
\author{Jung-Min Suh}
\email[E-mail: ]{suhjungmin@inha.edu}
\affiliation{Department of Physics, Inha University, Incheon 22212,
Republic of Korea}
\author{June-Young Kim}
\email[E-mail: ]{Jun-Young.Kim@ruhr-uni-bochum.de}
\affiliation{Ruhr-Universit\"at Bochum, Fakult\"at f\"ur Physik und
  Astronomie, Institut f\"ur Theoretische Physik II, D-44780 Bochum,
  Germany} 
\author{Ghil-Seok Yang}
\email[E-mail: ]{ghsyang@hoseo.edu}
\affiliation{Department of General Education for Human Creativity,
  Hoseo University, Asan 31499, 
Republic of Korea}
\author{Hyun-Chul Kim}
\email[E-mail: ]{hchkim@inha.ac.kr}
\affiliation{Department of Physics, Inha University, Incheon 22212,
Republic of Korea}
\affiliation{School of Physics, Korea Institute for Advanced Study
(KIAS), Seoul 02455, Republic of Korea}

\date{\today}
\begin{abstract}
We investigate the quark spin content of the SU(3) baryon octet,
decuplet, antitriplet, and sextet in a pion mean-field approach or the
chiral quark-soliton model, considering the $1/N_c$ rotational
corrections and SU(3) symmetry breaking effects. We compare the
present results with those from various theoretical works and lattice
QCD. 
\end{abstract}
\maketitle
\section{Introduction}
Understanding the spin structure of the nucleon has been one of the
most intriguing issues well over the decades. In quantum
chromodynamics~(QCD), the spin of the nucleon is known to be made of
the intrinsic spin and orbital angular momentum of its partons~(quarks
and gluons). It attracts considerable attention and becomes one of the
most important missions of the upcoming Electron-Ion Collider~(EIC)
project. A series of experiments on the spin structure
has been carried out by measuring the spin asymmetry in the polarized
lepton-nucleon deep inelastic
scattering~(DIS)~\cite{EuropeanMuon:1987isl, EuropeanMuon:1989yki,  
  SpinMuon:1993gcv, E142:1993hql, Ellis:1993te, 
  SpinMuonSMC:1995bzf, SpinMuon:1997yns} since the 1980s and has been
renewed by consecutive experiments~\cite{HERMES:2003gbu,
  HERMES:2004zsh, COMPASS:2006mhr, COMPASS:2010wkz}. In these
experiments, the spin structure function $g_{1}$ of 
the nucleon was extensively measured. The first moment of the
structure function $g_{1}$ is related to the singlet axial charge 
$g^{(0)}_{A}$. The experimental results imply that 
only a small fraction of the nucleon spin is carried by 
quarks. Thus, the Ellis-Jaffe sum rule~\cite{Ellis:1973kp}, 
i.e. $\Delta s =\Delta \bar{s} = 0$, does not hold anymore. 
It means that the sea quarks play a crucial role 
in understanding the spin structure of the nucleon~($\Delta s \sim
-0.10$~\cite{Bass:2004xa}). To explore the antiquark flavor asymmetry, 
the longitudinally polarized semi-inclusive deep inelastic
scattering~(SIDIS) has been also considered, which enables one to
measure the sea quark contributions isolated from the valence quark
ones. In addition to the sea quark contributions, one of the
ambiguities arises from the QCD axial anomaly
$\mathrm{U}(1)_{\mathrm{A}}$. In contrast to the traditional parton 
model, the gluon spin contributions are mixed with the quark one, so
that they cannot be treated independently. The
values of the isotriplet and octet axial charges, $g^{(3)}_{A}=1.2754 \pm
0.0013$~\cite{Workman:2020zbs} and $g^{(8)}_{A}=0.58 \pm
0.03$~\cite{Close:1993mv}, are respectively extracted from the neutron
$\beta$-decay and the hyperon semileptonic decays (HSDs) with SU(3)
symmetry imposed. There are recent
studies on the isovector axial charge~\cite{Alvarado:2021ibw,
  Walker-Loud:2019cif, Yao:2017fym, Liang:2016fgy, Capitani:2012gj} 
and flavor octet charge~\cite{Lorce:2007as, Bass:2009ed}. Note that
the value of the $g^{(8)}_{A}$ is still under debate. One also has to
keep in mind that while they are scale-independent, the isosinglet 
axial charge depends on the renormalization scale. 
With knowledge of the isotriplet and octet axial charges, the
scale-dependent $g^{(0)}_{A}$ was estimated from the
structure function $g_{1}$. The widely accepted value of the singlet
axial charge is given as $g^{(0)}_{A} \sim
0.33$~\cite{Aidala:2012mv}. 
There have been extensive studies on the axial charge of the baryon
octet: for example, chiral perturbation theory~\cite{Jiang:2009sf},
the lattice QCD~\cite{Lin:2007ap, Alexandrou:2017oeh,
  Berkowitz:2017gql, Liang:2018pis, Lin:2018obj}, a global QCD 
analysis~\cite{deFlorian:2009vb}, NNPDF collaboration
data~\cite{Nocera:2014gqa}, and COMPASS
data~\cite{COMPASS:2015mhb}. The axial charges of the $\Delta$ 
baryon were also estimated in chiral perturbation
theory~\cite{Jenkins:1991es,Jiang:2008we}, and those of the baryon
octet and decuplet were investigated in the relativistic constituent
quark  model (RCQM)~\cite{Choi:2010ty, Choi:2013ysa} and the
perturbative chiral quark model(PCQM)~\cite{Liu:2018jiu}. Recently,
the axial charges of the hyperons and charmed baryons were also
obtained from lattice QCD~\cite{Alexandrou:2016xok}. 

In the present work, we will investigate the quark spin content of the
SU(3) light and singly-heavy baryons within the frame of the chiral
quark-soliton model~($\chi$QSM) or pion-mean field approach. 
The model is based on Witten's seminal idea~\cite{Witten:1979kh,
  Witten:1983tw}. He proposed that a light baryon can be viewed as
$N_{c}$ valence quarks bound by the pion mean field in the large
$N_{c}$ limit. The $\chi$QSM was developed, based on the 
effective chiral action that was inspired by the QCD instanton
vacuum~\cite{Diakonov:1983hh, Diakonov:2002fq}. The axial charges of
the nucleon have been studied in both the
SU(2)~\cite{Wakamatsu:1993nq, Christov:1993ny, Christov:1995vm, 
  Praszalowicz:1995vi, Praszalowicz:1998jm} and SU(3)
$\chi$QSMs~\cite{Blotz:1993dd, Blotz:1994wi, Christov:1995vm,
  Praszalowicz:1998jm, Kim:1999uf, Wakamatsu:2002kq, Wakamatsu:2003wg,
  Silva:2005fa, Jun:2020lfx}. The value of the axial charge
$g^{(0)}_{A}$ derived from the $\chi$QSM is comparable to that 
extracted from the experiment. Moreover, the polarization of the
strange quark inside the nucleon, $\Delta s$ is in good agreement with
the data. 
The light-cone quantization of the $\chi$QSM having been employed, the
same quantities were examined~\cite{Diakonov:2004as,
  Lorce:2007as}. Since there are no gluon degrees of freedom in the
$\chi$QSM, the singlet axial charges become the  
scale dependence. However, it does not mean $g^{(0)}_{A}=1$, since the
missing contributions to the nucleon spin are attributed to the
orbital angular motion of the quarks~\cite{Wakamatsu:1990ud,
  Wakamatsu:1999nj, Ossmann:2004bp, Goeke:2007fp, Lorce:2011ni, Won}. 

While the quark spin content inside the light baryon has been
extensively investigated on both the theoretical and experimental 
sides, only a few works on the spin content of the singly heavy
baryons have been carried out~\cite{Alexandrou:2016xok}. Recently, the
$\chi$QSM was successfully extended to singly heavy baryons.  
In the heavy quark mass limit, i.e., $m_{Q}\to \infty$, a singly
heavy baryon is considered as $N_{c}-1$ valence quarks bound by the
pion mean field. Here, the heavy quark can be taken to be 
a mere static color source. Accordingly, the allowed SU(3) 
representations of the lowest-lying colored soliton with spin $J$ are
correctly identified as the antitriplet~$(\overline{\bm{3}})$ with
spin $J=0$ and sextet~$({\bm{6}})$ with spin $J=1$. The color-singlet
heavy baryon is then constructed by coupling a heavy quark with spin
$J_{Q}=1/2$. As a result, the heavy baryon with spin $J'$ form one 
antitriplet~$(\overline{\bm{3}}, J'=1/2)$ and two sextets~$(\bm{6},
J'=1/2)$ and ~$(\bm{6}, J'=3/2)$. The two sextet multiplets are
subjected to a hyperfine splitting. Based on the modified pion 
mean field and the heavy-quark symmetry, various properties of the
singly heavy baryons have been described well such as the mass
splittings~\cite{Yang:2016qdz, Kim:2018xlc, Kim:2019rcx},
electromagnetic properties~\cite{Yang:2018uoj, Kim:2018nqf,
  Yang:2019tst, Kim:2019wbg, Kim:2020uqo, Kim:2021xpp}, strong
decays~\cite{Kim:2017khv, Suh:2022ean} , and gravitational form
factors~\cite{Kim:2020nug}. Thus, the $\chi$QSM allows one to
investigate both the light and singly heavy baryons on an equal
footing. In the current work, we want to scrutinize the axial charges 
of both the light and singly heavy baryons with spin-1/2 and -3/2. We
will then discuss the results for the singly heavy baryons compared
with those for light baryons.  

The present work is organized as follows: In Sec~\ref{sec:2}, we
describe the formalism for the axial charges of both the light and
singly heavy baryons within the $\chi$QSM, considering the
rotational $1/N_c$ corrections and the effects of flavor SU(3)
symmetry breaking. In Sec~\ref{sec:3}, results for the axial charges
and spin contents of the both light and singly heavy baryons are
discussed. We summarize the current work and draw
conclusions in the last section.    

\section{Axial charges in the $\chi$QSM \label{sec:2}}
The axial charges for both the spin-1/2 and -3/2 particles are
respectively given by the matrix element of the axial-vector current
$A^{(\chi)}_{\mu}=\bar{\psi}\gamma_{\mu}\gamma^{5}\lambda^{\chi}\psi +
\bar{\Psi} \gamma_{\mu} \gamma^{5} \Psi$: 
\begin{align}
\langle B (p,J'_{3})| A^{(\chi)}_{\mu}  | B (p,J'_{3}) \rangle & =
  g^{(\chi)}_{A} \bar{u}(p,J'_{3}) \gamma_{\mu} \gamma_{5}
  {u}(p,J'_{3}), \cr
\langle B (p,J'_{3})| A^{(\chi)}_{\mu}  | B (p,J'_{3}) \rangle &=
  -g^{{(\chi)}}_{A}\bar{u}^{\alpha}(p,J'_{3})  \gamma_{\mu} \gamma_{5}
  u_{\alpha}(p,J'_{3}), 
\end{align}
where $p$ and $J'_{3}~(J_{3})$ denote the momentum and the spin
polarization of the heavy~(light) baryon $B$,
respectively. The flavor index runs over $\chi =0,3,8$, and 
$\lambda^{\chi}$ stands for the well-known SU(3) Gell-Mann
matrices. We define $\lambda^{0}$ as $3\times3$ unit matrix in the
flavor space. The $\psi=(u,d,s)$ and $\Psi$ designate the light-quark
and heavy-quark field operators, respectively. Here, one should keep
in mind that the heavy quark flavor represents both the charm and
bottom quarks. Assuming the heavy quark spin symmetry, we have the
degenerate axial charges of the charmed and bottom baryons due to the
heavy-quark flavor symmetry. Thus, we consider the singly charmed
baryons in the present work.  $u(p)$ and $u^{\alpha}(p)$ are
respectively the Dirac spinor and the Rarita-Schwinger spinor   
that satisfies the subsidiary conditions $\gamma^{\alpha}
u_{\alpha}(p) = p^{\alpha}u_{\alpha}(p)=0$~\cite{Rarita:1941mf}. These
definitions of the axial charges are related to the first moment of
the longitudinally polarized quark distributions $g_{1}$ accessed by
the DIS experiment:  
\begin{align}
g^{(0)}_{A} &= \sum_{q=u,d,s,c} \Delta q = \Delta \Sigma, \ \ \  \ \ \
   g^{(3)}_{A} = \Delta u - \Delta d, \ \ \ \ \ \
   g^{(8)}_{A} = \frac{1}{\sqrt{3}}\left(\Delta u + \Delta
   d - 2 \Delta s\right). 
\end{align}
Here, one has to bear in mind that there are no gluonic degrees of
freedom in the $\chi$QSM as mentioned previously, so that we drop the
gluon contribution to $g^{(0)}_{A}$ induced by the U(1) axial anomaly.  

In the light-baryon sector, the matrix element of the axial-vector
current for the light baryon has been considered in
Ref.~\cite{Suh:2022guw, Suh:2022ean, Jun:2020lfx}. All dynamical
information on the axial-vector properties has been expressed in
terms of the collective operators together with the dynamical
parameters. They are written by the six dynamical parameters
$a_{1\ldots6}$ with the explicit flavor SU(3) symmetry breaking
considered. They are reduced to the three dynamical parameters 
$a_{1,2,3}$ with flavor SU(3) symmetry. The general expressions of the
collective operators for the axial-vector charges have been
constructed already in previous works~\cite{Kim:1995ha, Kim:1997ip,
  Yang:2015era, Kim:1995mr}:   
\begin{align}
\hat{g}_{A}^{(\chi)} &=a_{1}D_{\chi 3}^{(8)}+a_{2}d_{pq3}D_{\chi p}^{(8)}\hat{J}_{q}
+\frac{a_{3}}{\sqrt{3}}D_{\chi 8}^{(8)}\hat{J}_{3}
+\frac{a_{4}}{\sqrt{3}}d_{pq3}D_{\chi p}^{(8)}D_{8 q}^{(8)}
+a_{5}(D_{\chi 3}^{(8)}D_{88}^{(8)}+D_{\chi 8}^{(8)}D_{83}^{(8)}) \cr
&+a_{6}(D_{\chi 3}^{(8)}D_{88}^{(8)}-D_{\chi 8}^{(8)}D_{83}^{(8)}).
\label{eq:8}
\end{align}
where the indices of the SU(3) symmetric tensor $d_{pq3}$ run over
$p,q=4\ldots 7$. $\hat{J}_{3}$ and $\hat{J}_{q}$ denote the SU(3) spin
operators. $D^{(\nu)}_{a b}$ corresponds to the SU(3) Wigner matrices
with its representation $\nu$. The dynamical parameters $a_{1,2,3}$
are determined from the model calculation. Their explicit expressions
can be found in Appendix~\ref{App:a}. One great merit of the $\chi$QSM
is that it can describe both the light and singly heavy baryons within
the same pion mean-field approximation. Recently, the $\chi$QSM was
extended to the heavy baryon sector~\cite{Suh:2022guw, Suh:2022ean,
  Jun:2020lfx}. To describe the singly heavy baryons, the pion mean
fields created by the $N_c$ valence quarks are replaced by those with
$N_c-1$ quarks, because the heavy quark inside a singly heavy baryon
can be stripped off. The structure of the collective operators is
left to be the same. Thus, there is nothing changed from the
light-baryon collective operators in Eq.~\eqref{eq:8} except for the
values of the dynamical parameters of which both the light and heavy
baryons are listed in the following subsections. 

To evaluate the matrix elements of the collective operators, the
collective wave functions of both the light and singly heavy baryons
should be provided. With $N_Q$ defined as the number of heavy quarks,
the wave function of the light-quark state with flavor state $(Y , T ,
T_{3})$ and spin state $(Y' =-(N_{c}-N_{Q})/3, J, J_{3})$ in the
flavor SU(3) representation $\nu$ is given by
$\psi^{(\nu)}_{(Y,T,T_{3}) (Y',J,J_{3})}$ in terms of the Wigner
rotational $D(R)$ matrices:  
\begin{align}
| B \rangle := \psi^{(\nu)}_{(Y,T,T_{3}) (Y',J,J_{3})}(R)
  =\sqrt{\mathrm{dim}(\nu)}(-1)^{\mathcal{Q}_s}[D_{(Y,T,T_{3})
  (-Y',J,-J_{3})}^{(\nu)}(R)]^{*}, 
\label{eq:6}
\end{align}
where dim($\nu$) denotes the dimension of the representation $\nu$,
and $\mathcal{Q}_s$ stands for a charge corresponding to the baryon
state $S$, i.e. $\mathcal{Q}_s=J_{3}+Y'/2$. If one takes $N_{Q}=0$,
the right hypercharge becomes $Y_{R}=-Y'=N_{c}/3$, which corresponds
to the well-known light-baryon collective wave function. Furthermore,
to construct the heavy-baryon collective wave function, we need to
couple this light-quark state to the heavy-quark spinor with the SU(2) 
Clebsch-Gordan coefficient
$C_{J,J_{3},J_{Q},J_{Q_3}}^{J',J'_{3}}$. The heavy-baryon collective 
wave function is then constructed as: 
\begin{align}
| B_{Q}  \rangle := \psi_{B_{Q}}^{(\nu)}(R)
  =\sum_{J_{3},J_{Q_3}}C_{J,J_{3},J_{Q},J_{Q_3}}^{J',J'_{3}} 
\chi_{J_{Q_3}}\psi^{(\nu)}_{(Y,T,T_{3})(Y',J,J_{3})}(R),
\label{eq:7}
\end{align}
where $\chi_{J_{Q_3}}$ represents the Pauli spinors. By taking $N_{Q}=1$,
the right hypercharge becomes $Y_{R}=-Y'=(N_{c}-1)/3$, which
corresponds to the singly heavy-baryon collective wave
function. Though one can construct the doubly heavy-baryon wave
function in the same manner, the corresponding pion mean field is not
strong enough to construct the stable soliton~\cite{Kim:2019rcx}. Thus,
in this work, we will concentrate on the axial charges 
of the light~$(N_{Q}=0)$ and singly heavy~$(N_{Q}=1)$ baryons. The
axial charges for both the light and heavy baryons can be 
obtained by sandwiching the collective operators between their
collective wave functions as follow: 
\begin{align}
g^{(\chi),B(B_{Q})}_{A}= \langle B (B_{Q}) | \hat{g}^{(\chi)}_{A} | B
  (B_{Q}) \rangle. 
\label{eq:sand}
\end{align}

\subsection{Axial charges of the baryon octet and decuplet}
All the values of the dynamical parameters for the axial properties in
the light-baryon sector are obtained in Refs.~\cite{Suh:2022guw,
  Jun:2020lfx}:  
\begin{align}
 a_{1}=-3.74, \quad  a_{2}=2.34, \quad a_{3}=0.88.
\label{eq:Light_v}
\end{align}
In the exact SU(3) symmetry, baryonic axial-vector amplitudes are
given in terms of two reduced matrix elements $F$ and $D$, which is
drawn from the SU(3) Wigner-Eckart theorem. Since these values $F$ and 
$D$ are broadly used, we express our dynamical parameters $a_{1,2,3}$
in terms of them: 
\begin{align}
&F=-\frac{1}{12}\left(a_{1}-\frac{1}{2}a_{2}\right)+\frac{1}{24}a_{3}
  = 0.446, \quad 
D=-\frac{3}{20}\left(a_{1}-\frac{1}{2}a_{2}\right)-\frac{1}{40}a_{3}
  =0.716, 
\label{eq:9}
\end{align}
which is consistent with the results for the $F$ and $D$ presented in
Ref.~\cite{Ledwig:2008ku}. 

These results are in agreement with
the empirical ones extracted from the
experiments~\cite{AsymmetryAnalysis:1999gsr}, $F= 0.463 \pm 0.008$ and 
$D = 0.804 \pm 0.008$. There are two different lowest-lying
representations of the light baryons: the baryon octet $({\bm{8}},
J=1/2)$ and the baryon decuplet $({\bm{10}}, J=3/2)$. From
Eqs.~\eqref{eq:sand} and \eqref{eq:Light_v}, we first derive the 
expressions for the axial charges of the baryon octet in the exact 
SU(3) symmetry in terms of $F$ and $D$ for the flavor-singlet 
\begin{align}
&\hat{g}^{(0),B_{8}}_{A} =9F-5D = 0.44, 
\end{align}
and for the flavor-triplet
\begin{align}
&\hat{g}^{(3),N}_{A} =2T_{3}\left(F+D\right), \quad
  \hat{g}^{(3),\Lambda}_{A} =0, \quad \hat{g}^{(3),\Sigma}_{A} =2
  T_{3}F, \quad \hat{g}^{(3),\Xi}_{A} =2T_{3}(F-D), 
\end{align}
and for the flavor-octet
\begin{align}
&\hat{g}^{(8),N}_{A} = \frac{\sqrt{3}}{3}(3F-D), \quad
  \hat{g}^{(8),\Lambda}_{A} =-\frac{2\sqrt{3}}{3}D, \quad
  \hat{g}^{(8),\Sigma}_{A} = \frac{2\sqrt{3}}{3}D, \quad
  \hat{g}^{(8),\Xi}_{A} = -\frac{\sqrt{3}}{3}(3F+D). 
\end{align}
In the same manner, we derive the axial charges of the baryon decuplet:
\begin{align}
&\hat{g}^{(0),B_{10}}_{A} = 9F-5D,  \quad \hat{g}^{(3),B_{10}}_{A} =
  T_{3}F, \quad \hat{g}^{(8),B_{10}}_{A} = \frac{\sqrt{3}Y}{2}F, 
\label{eq:C5}
\end{align}
where we set all the spin polarization of the baryons to be
$J_{3}=1/2$ from now on. As discussed in
Ref.~\cite{Blotz:1993dd, Blotz:1994wi, Praszalowicz:1998jm,
  Kim:2017khv}, the dynamical parameters can be expressed in terms
of the moments of inertia ($I_{1}$ and $I_{2}$) and have their own 
$N_{c}$ factors. While $a_{1}$ consists of both the $N_{c}$ and
$N^{0}_{c}$ order terms, $a_{2}$ and $a_{3}$ are of order 
$N^{0}_{c}$ (see Appendix~\ref{App:a}). Such a relation can be clearly
seen in the non-relativistic~(NR) limit, i.e., the small soliton size
limit. The dynamical parameters in this limit are found to be 
\begin{align}
a_{1} \xrightarrow[\text{NR}]{} -(N_{c}+2), \quad
a_{2} \xrightarrow[\text{NR}]{} 4, \quad
a_{3} \xrightarrow[\text{NR}]{} 2. 
\label{eq:11}
\end{align}
In this limit, we recover the results from the naive quark model for the 
$p~(J_{3}=1/2)$ and $\Delta^{+}~(J_{3}=1/2)$ axial charges 
\begin{align}
&g^{(0),p}_{A} \xrightarrow[\text{NR}]{} 1, \quad g^{(3),p}_{A}
  \xrightarrow[\text{NR}]{} \frac{5}{3}, \quad g^{(8),p}_{A}
  \xrightarrow[\text{NR}]{} \frac{1}{\sqrt{3}}, 
\end{align}
and
\begin{align}
&g^{(0),\Delta^{+}}_{A}  \xrightarrow[\text{NR}]{} 1, \quad
  g^{(3),\Delta^{+}}_{A}   \xrightarrow[\text{NR}]{} \frac{1}{3},
  \quad g^{(8),\Delta^{+}}_{A}  \xrightarrow[\text{NR}]{}
  \frac{1}{\sqrt{3}}. 
\end{align}

\subsection{Axial charges of singly heavy baryons}
In the presence of the $N_{c}-1$ valence quarks, the pion mean field 
becomes weaker than that in the light-baryon system, so that all the
values of the dynamical parameters related to the axial properties for
the singly heavy baryon are also changed. These dynamical parameters
are obtained as 
\begin{align}
 a_{1}=-3.19, \quad  a_{2}=2.98, \quad a_{3}=1.32
 \label{eq:dynamical_HB}
\end{align}
in the present model. Keeping the same definition used in 
Eq.~\eqref{eq:9}, we obtain $F=0.446$ and $D=0.671$ in the
heavy-baryon sector. While the value of $F$ is almost intact,
that of $D$ is slightly changed.  

There are three different lowest-lying representations of the singly
heavy baryons: the baryon antitriplet $(\overline{\bm{3}}, J=0,J_{Q}=1/2,
J'={1/2})$, the baryon sextet $({\bm{6}},J=1,J_{Q}=1/2,J'={1/2})$ with
spin 1/2 and the sextet $({\bm{6}}, J=1,J_{Q}=1/2, J'={3/2})$ with
spin 3/2. In the case of the antitriplet baryons, their spins are 
constructed out of light-quark pair with $J=0$ and the heavy quark
state with $J_Q=1/2$, which means that there is no light-quark
contribution to all the vector and axial-vector quantities, such 
as the magnetic moments and axial charges of the baryon
antitriplet. So, the quark spin contribution to the axial
charges of the antitriplet baryons only comes from the heavy
quark. Since the results for the baryon antitriplet are trivial, we 
will not deal with them in this work.  The light quarks contribute to
the axial charges for the baryon sextet, which dominate over the
heavy-quark contributions. The flavor-singlet, -triplet, and -octet
axial charges of the baryon sextet with $J'=1/2$ are given by  
\begin{align}
&\hat{g}^{(0),B_{6,J'=1/2}}_{A}= \frac{4}{3}(9F-5D)-\frac{1}{3}, \quad
  \hat{g}^{(3),B_{6,J'=1/2}}_{A} = \frac{T_{3}}{15}(27F+5D), \quad
  \hat{g}^{(8),B_{6,J'=1/2}}_{A}= \frac{\sqrt{3}Y}{30}(27F+5D).  
\label{eq:6J12sym}
\end{align}

Interestingly, the axial charges of the baryon sextet with $J'=3/2$
can be related to those of the baryon sextet with $J'=1/2$ as follows: 
\begin{align}
&\hat{g}^{(0),B_{6,J'=3/2}}_{A}=
  \frac{1}{2}\left(\hat{g}^{(0),B_{6,J'=1/2}}_{A} + 1\right), \quad
  \hat{g}^{(\chi=3,8),B_{6,J'=3/2}}_{A} =
  \frac{1}{2}\hat{g}^{(\chi=3,8),B_{6,J'=1/2}}_{A}. 
\label{eq:6J32sym}
\end{align}
We observe that the light-quark contributions to the axial charges of
the baryon sextet with $J'=1/2$ differ from those with
$J'=3/2$ only by factor 2. 

Being similar to the light-baryon sector, the $a_{1}$ consists of both
the $(N_{c}-1)$ and $N^{0}_{c}$ order terms, and $a_{2}$ and $a_{3}$
are of order $N^{0}_{c}$ (see Appendix~\ref{App:a}). In the
non-relativistic (NR) limit, $a_1$, $a_{2}$, and $a_{3}$ are reduced
to 
\begin{align}
a_{1} \xrightarrow[\text{NR}]{} -[(N_{c}-1)+2], \quad
a_{2} \xrightarrow[\text{NR}]{} 4, \quad
a_{3} \xrightarrow[\text{NR}]{} 2 .
\label{eq:12}
\end{align}
Using the values $a_{1,2,3}$ in the NR limit,
we show that the $\Sigma^{++}_{c}$ and $\Sigma^{*++}_{c}$ axial
charges are led to 
\begin{align}
g^{(0),\Sigma^{++}_{c}}_{A}  \xrightarrow[\text{NR}]{} 1, \quad
  g^{(3),\Sigma^{++}_{c}}_{A} \xrightarrow[\text{NR}]{} \frac{4}{3},
  \quad  
g^{(8),\Sigma^{++}_{c}}_{A} \xrightarrow[\text{NR}]{}
  \frac{4}{3\sqrt{3}},   
\end{align}
and 
\begin{align}
g^{(0),\Sigma^{*++}_{c}}_{A}  \xrightarrow[\text{NR}]{} 1, \quad
  g^{(3),\Sigma^{*++}_{c}}_{A} \xrightarrow[\text{NR}]{} \frac{2}{3},
  \quad  
g^{(8),\Sigma^{*++}_{c}}_{A} \xrightarrow[\text{NR}]{}
  \frac{2}{3\sqrt{3}}.  
\end{align}
Here the heavy quark makes the flavor-singlet axial charges shifted by 
$-1/3$ and $1/3$ in the baryon sextets with $J'=1/2$ and $J'=3/2$,
respectively. 

\subsection{Effects of flavor SU(3) symmetry breaking}
In this subsection, we will consider the effects of flavor SU(3)
symmetry breaking on the axial charges of both the light and singly
heavy baryons. As discussed already, the terms proportional to
$a_{4,5,6}$ arise from the perturbative treatment of the strange
current quark mass, so that the collective wave functions for the
soliton are no longer pure state but are admixed with higher
representations. Thus, to derive the collective wave functions, we
need to diagonalize the collective Hamiltonian with flavor SU(3)
symmetry breaking. The collective Hamiltonian is given by 
\begin{align}
H_{\mathrm{coll}} = H_{\mathrm{sym}} + H_{\mathrm{sb}},
\end{align}
where
\begin{align}
  \label{eq:Hamiltonian}
H_{\mathrm{sym}} = M_{\mathrm{cl}} + \frac1{2I_1} \sum_{i=1}^3
  \hat{J}_i^2 + \frac1{2I_2} \sum_{p=4}^7 \hat{J}_p^2,\quad
H_{\mathrm{sb}} = \alpha D_{88}^{(8)} + \beta \hat{Y} +
  \frac{\gamma}{\sqrt{3}} \sum_{i=1}^3 D_{8i}^{(8)} \hat{J}_i.
\end{align}
Here, $I_1$ and $I_2$ stand for the moments of inertia for the soliton,
 and $D^{(8)}_{ab}$ denotes SU(3) Wigner $D$ functions. 
The dynamical parameters $\alpha$, $\beta$, and $\gamma$, which arise from 
flavor SU(3) symmetry breaking, are expressed in terms of the
moments of inertia $I_1$ and $I_2$, and the anomalous moments of
inertia $K_1$ and $K_2$
\begin{align}
\alpha=\left (-\frac{{\Sigma}_{\pi N}}{3\overline{m}}
  +\frac{K_{2}}{I_{2}} \right )m_{\mathrm{s}},
  \quad \beta=-\frac{ K_{2}}{I_{2}}m_{\mathrm{s}}, 
  \quad \gamma=2\left ( \frac{K_{1}}{I_{1}}-\frac{K_{2}}{I_{2}}
  \right ) m_{\mathrm{s}},
\label{eq:alphaetc}  
\end{align}
where $\Sigma_{\pi N}$ stands for the pion-nucleon $\Sigma$ term. 
By diagonalizing the collective Hamiltonian with flavor SU(3) symmetry
breaking and by setting $N_{Q}=0$ in the dynamical parameters listed
in Appendix~\ref{App:a}, we obtain the following wave functions for
baryon octet and decuplet  
\begin{align}
&|B_{{\bm8}}\rangle = |{\bm{8}},B\rangle + 
  c^{B}_{{\overline{10}}}|{{\bm{\overline{10}}}},B\rangle + 
  c^{B}_{{27}}|{{\bm{27}}},B\rangle, \cr
&|B_{{\bm{10}}_{3/2}}\rangle = |{\bm{10}}_{3/2},B\rangle + 
  a^{B}_{{27}}|{{\bm{27}}}_{3/2},B\rangle + 
  a^{B}_{{35}}|{{\bm{35}}}_{3/2},B\rangle,
\label{eq:mixedWF1}
\end{align}
with the mixing coefficients
\begin{eqnarray}
c_{{\overline{10}}}^{B}
\;=\;
c_{{\overline{10}}}\left[\begin{array}{c}
\sqrt{5}\\
0 \\
\sqrt{5} \\
0
\end{array}\right], 
\quad
c_{27}^{B}
\;=\; 
c_{27}\left[\begin{array}{c}
\sqrt{6}\\
3 \\
2 \\
\sqrt{6}
\end{array}\right], 
\quad
a_{{27}}^{B}
\;=\;
a_{{27}}\left[\begin{array}{c}
\sqrt{15/2}\\
2 \\
\sqrt{3/2} \\
0
\end{array}\right], 
\quad
a_{35}^{B}
\;=\; 
a_{35}\left[\begin{array}{c}
5/\sqrt{14}\\
2\sqrt{{5}/{7}} \\
3\sqrt{{5}/{14}} \\
2\sqrt{{5}/{7}}
\end{array}\right],
\label{eq:pqmix}
\end{eqnarray}
respectively, in the basis $\left[N,\;\Lambda,\;\Sigma,\;\Xi\right]$
for the octet and $\left[\Delta,\;\Sigma^{*},\;\Xi^{*},\;\Omega\right]$
for the decuplet. The coefficients $c_{{\overline{10}}}$, $c_{27}$, 
$a_{27}$, and $c_{35}$ are written as 
\begin{eqnarray}
c_{{\overline{10}}} \;=\;
{\displaystyle -\frac{{I}_{2}}{15} \left ( \alpha + \frac{1}{2}
  \gamma \right)}, \quad 
c_{27} \;=\; {\displaystyle -\frac{{I}_{2}}{25} \left( \alpha 
  -\frac{1}{6}\gamma \right)}, \quad
  a_{27} \;=\;
{\displaystyle -\frac{{I}_{2}}{8} \left ( \alpha + \frac{5}{6}
  \gamma \right)}, \quad
a_{35} \;=\; {\displaystyle -\frac{{I}_{2}}{24} \left( \alpha 
  -\frac{1}{2}\gamma \right)}. 
\label{eq:pqmix2}
\end{eqnarray}

By setting $N_{Q}=1$ in the dynamical parameters listed in
Appendix~\ref{App:a}, the collective wave functions for the baryon
sextets are obtained as  
\begin{align}
|B_{{\bm{6}}_{1}}\rangle &= |{\bm{6}}_{1},B\rangle 
  + q_{\overline{15}}^{B}|\bm{\overline{15}}_{1},B\rangle 
  + q_{\overline{24}}^{B}|\bm{\overline{24}}_{1},B\rangle,
\label{eq:mixedWF2}
\end{align}
with the mixing coefficients
\begin{eqnarray}
q_{\overline{15}}^{B}
\;=\; 
q_{\overline{15}}\left[\begin{array}{c}
\sqrt{5}/5\\
\sqrt{30}/20 \\
0
\end{array}\right], \quad
q_{\overline{24}}^{B}
\;=\;
q_{\overline{24}}\left[\begin{array}{c}
-\sqrt{10}/10\\
-\sqrt{15}/10 \\
-\sqrt{15}/10
\end{array}\right], 
\label{eq:pqmix1}
\end{eqnarray}
respectively, in the basis of
$\left[\Sigma_{c}(\Sigma_{c}^{*}),\;\Xi_{c}^{\prime}(\Xi_{c}^{*}) 
,\;\Omega_{c}^{0}(\Omega_{c}^{*0})\right]$ for the baryon sextets. 
The coefficients $q_{\overline{15}}$ and $q_{\overline{24}}$ are given
in terms of the dynamical parameters 
$\alpha$ and $\gamma$
\begin{align}
q_{\overline{15}} = -\frac{1}{\sqrt{2}} \left( \alpha +
  \frac{2}{3}\gamma \right)I_{2},\quad  
q_{\overline{24}} = \frac{4}{5\sqrt{10}} \left( \alpha -
  \frac{1}{3}\gamma \right)I_{2}. 
\label{eq:pqmix2}
\end{align}
The numerical results for the dynamical parameters are found to be 
\begin{align}
 a_{1}=-3.75, \quad  a_{2}=2.34, \quad a_{3}=0.88, \quad a_{4}=-0.29,
  \quad a_{5}=-0.01, \quad a_{6}=0.02 
\end{align}
for the light baryons, and 
\begin{align}
 a_{1}=-3.20, \quad  a_{2}=2.98, \quad a_{3}=1.32, \quad a_{4}=-0.23,
  \quad a_{5}=-0.02, \quad a_{6}=0.03 
\end{align}
for the singly heavy baryons. Note that values of the $a_{1}$ are
slightly different from those with flavor SU(3) symmetry in
Eqs.~\eqref{eq:Light_v} and~\eqref{eq:dynamical_HB}. As discussed in
the previous subsections, 
the leading contributions to the axial charges are expressed in
terms of $F$ and $D$ or $a_{1,2,3}$. We now introduce the contributions
from the flavor SU(3) symmetry breaking. They consist of two
different terms, i.e., that from the effective chiral action and that
from the collective wave functions. Here, we will refer to them
respectively as ``operator corrections'' and ``wave function
corrections'' and denote them by the superscripts (op) and (wf). For
convenience, we introduce the following combinations of the 
dynamical parameters from the effective chiral action $(x',y',z')$, 
and from the collective wave functions (octet $p',q'$; decuplet $s',
n'$; sextets $l', m'$): 
\begin{align}
&x'=a_{4}, \quad
y'=a_{5}, \quad
z'=a_{6}, \cr
&p'=c_{\overline{10}}(a_{1}+a_{2}+\frac{1}{2}a_{3}), \quad
q'=c_{27}(a_{1}+2a_{2}-\frac{3}{2}a_{3}),\cr
&s'=a_{27}(2a_{1}+a_{2}+3a_{3}), \quad
n'=a_{35}(2a_{1}+5a_{2}-5a_{3}), \cr
&l'=q_{\overline{15}}(2a_{1}+a_{2}+2a_{3}), \quad
m'=q_{\overline{24}}(a_{1}+2a_{2}-2a_{3}).
\label{eq:9_2}
\end{align}
For the singlet axial charges, the expressions of the operator
corrections to them for the baryons octet and decuplet are found to
be: 
\begin{align}
&(\hat{g}^{(0),N}_{A})^{(\mathrm{op})} =-\frac{1}{5}(y'-z'), \cr 
&(\hat{g}^{(0),\Lambda}_{A})^{(\mathrm{op})} =\frac{3}{5}(y'-z'), \cr 
&(\hat{g}^{(0),\Sigma}_{A})^{(\mathrm{op})} =-\frac{3}{5}(y'-z'), \cr 
&(\hat{g}^{(0),\Xi}_{A})^{(\mathrm{op})} =\frac{4}{5}(y'-z'), \cr
&(\hat{g}^{(0),B_{10}}_{A})^{(\mathrm{op})} = -\frac{Y}{4}(y'-z').
\end{align}
For the wave function corrections to the singlet axial charges, they
vanish, because the inner product between the states with the
different representations is not allowed. 
\begin{align}
(\hat{g}^{(0),B_{8}}_{A})^{(\mathrm{wf})} =0, \quad
  (\hat{g}^{(0),B_{10}}_{A})^{(\mathrm{wf})} =0. 
\end{align}
For the triplet axial charges, the expressions for the baryons octet
and decuplet from all the operator corrections are proportional to  
$T_{3}$: 
\begin{align}
&(\hat{g}^{(3),N}_{A})^{(\mathrm{op})} =T_{3}\left(-\frac{11}{135}x' 
-\frac{2}{9}y'
-\frac{2}{15}z' \right),\cr 
&(\hat{g}^{(3),\Lambda}_{A})^{(\mathrm{op})} =0, \cr
&(\hat{g}^{(3),\Sigma}_{A})^{(\mathrm{op})} =T_{3}\left(-\frac{1}{30}x'
-\frac{1}{15}y'
+\frac{1}{15}z'\right), \cr
&(\hat{g}^{(3),\Xi}_{A})^{(\mathrm{op})} =T_{3}\left(\frac{2}{135}x'
+\frac{4}{45}y'
-\frac{2}{15}z'\right), \cr
&(\hat{g}^{(3),\Delta}_{A})^{(\mathrm{op})} =
  T_{3}\left(-\frac{11}{756}x'-\frac{5}{126}y'\right), \cr  
&(\hat{g}^{(3),\Sigma^{*}}_{A})^{(\mathrm{op})} =
  T_{3}\left(-\frac{5}{252}x'-\frac{1}{42}y'\right), \cr 
&(\hat{g}^{(3),\Xi^{*}}_{A})^{(\mathrm{op})} =
  T_{3}\left(-\frac{19}{756}x'-\frac{1}{126}y'\right), \cr 
&(\hat{g}^{(3),\Omega}_{A})^{(\mathrm{op})} = 0.
\end{align}
and those of the wave function corrections are also found to be
\begin{align}
&(\hat{g}^{(3),N}_{A})^{(\mathrm{wf})} = T_{3} \left(-\frac{2}{3}p'
-\frac{4}{45}q'\right), \cr 
&(\hat{g}^{(3),\Lambda}_{A})^{(\mathrm{wf})} =0, \cr
&(\hat{g}^{(3),\Sigma}_{A})^{(\mathrm{wf})} = -\frac{T_{3}}{3}p', \cr
&(\hat{g}^{(3),\Xi}_{A})^{(\mathrm{wf})} = \frac{4T_{3}}{45}q', \cr
&(\hat{g}^{(3),\Delta}_{A})^{(\mathrm{wf})} = T_{3}\left(-\frac{5}{72}m'
-\frac{1}{168}n'\right), \cr 
&(\hat{g}^{(3),\Sigma^{*}}_{A})^{(\mathrm{wf})} = T_{3}\left(-\frac{1}{12}m'
 -\frac{1}{84}n'\right),\cr
&(\hat{g}^{(3),\Xi^{*}}_{A})^{(\mathrm{wf})} = T_{3} \left(\frac{7}{72}m'
-\frac{1}{56}n' \right), \cr
&(\hat{g}^{(3),\Omega}_{A})^{(\mathrm{wf})} = 0.
\end{align}
Lastly, the expressions for the operator corrections to the octet axial
charges are given by 
\begin{align}
&(\hat{g}^{(8),N}_{A})^{(\mathrm{op})} = \frac{1}{30\sqrt{3}}x', \cr 
&(\hat{g}^{(8),\Lambda}_{A})^{(\mathrm{op})} = -\frac{1}{10\sqrt{3}}x', \cr
&(\hat{g}^{(8),\Sigma}_{A})^{(\mathrm{op})} =\frac{1}{18\sqrt{3}}x'
+\frac{2}{15\sqrt{3}}y', \cr
&(\hat{g}^{(8),\Xi}_{A})^{(\mathrm{op})} = -\frac{1}{15\sqrt{3}}x'
-\frac{1}{5\sqrt{3}}y', \cr
&(\hat{g}^{(8),\Delta}_{A})^{(\mathrm{op})} =
  \frac{5}{168\sqrt{3}}x'+\frac{1}{28\sqrt{3}}y', \cr  
&(\hat{g}^{(8),\Sigma^{*}}_{A})^{(\mathrm{op})} =
  -\frac{1}{126\sqrt{3}}x'+\frac{1}{42\sqrt{3}}y',\cr 
&(\hat{g}^{(8),\Xi^{*}}_{A})^{(\mathrm{op})} =
  -\frac{5}{168\sqrt{3}}x'-\frac{1}{28\sqrt{3}}y', \cr 
&(\hat{g}^{(8),\Omega}_{A})^{(\mathrm{op})} =
  -\frac{1}{28\sqrt{3}}x'-\frac{1}{7\sqrt{3}}y', 
\end{align}
and those of the wave function corrections are written as 
\begin{align}
&(\hat{g}^{(8)}_{A})^{(\mathrm{wf})}(N) = \frac{1}{\sqrt{3}}p',
-\frac{2}{5\sqrt{3}}q' \cr 
&(\hat{g}^{(8),\Lambda}_{A})^{(\mathrm{wf})} = -\frac{\sqrt{3}}{5}q', \cr
&(\hat{g}^{(8),\Sigma}_{A})^{(\mathrm{wf})} = \frac{1}{\sqrt{3}}p',
-\frac{4}{15\sqrt{3}}q' \cr
&(\hat{g}^{(8),\Xi}_{A})^{(\mathrm{wf})} = -\frac{2}{5\sqrt{3}}q', \cr
&(\hat{g}^{(8),\Delta}_{A})^{(\mathrm{wf})} = \frac{5}{16\sqrt{3}}s',
-\frac{5}{112\sqrt{3}}n', \cr 
&(\hat{g}^{(8),\Sigma^{*}}_{A})^{(\mathrm{wf})} = \frac{1}{6\sqrt{3}}s'
-\frac{1}{14\sqrt{3}}n',\cr
&(\hat{g}^{(8),\Xi^{*}}_{A})^{(\mathrm{wf})} = \frac{1}{16\sqrt{3}}s'
-\frac{3\sqrt{3}}{112}n', \cr
&(\hat{g}^{(8),\Omega}_{A})^{(\mathrm{wf})} = -\frac{1}{14\sqrt{3}}n'.
\end{align}

WHen it comes to the singly heavy baryons, we also obtain
the expressions for both the operator and wave function corrections to
the singlet axial charges 
\begin{align}
&(\hat{g}^{(0),B_{6,J=1/2}}_{A})^{(\mathrm{op})} =
  -\frac{3Y}{5}(y'-z'), \quad
  (\hat{g}^{(0),B_{6,J=1/2}}_{A})^{(\mathrm{wf})} = 0, 
\end{align}
and the triplet axial charges 
\begin{align}
&(\hat{g}^{(3),\Sigma_{c}}_{A})^{(\mathrm{op})}= T_{3}\left(-\frac{1}{27}x'
-\frac{4}{45}y'\right), \cr
&(\hat{g}^{(3),\Xi_{c}}_{A})^{(\mathrm{op})}= T_{3} \left(-\frac{14}{27}x'
-\frac{2}{45}y'\right), \cr 
&(\hat{g}^{(3),\Omega_{c}}_{A})^{(\mathrm{op})}= 0, \cr
&(\hat{g}^{(3),\Sigma_{c}}_{A})^{(\mathrm{wf})}=
  T_{3}\left(-\frac{2}{9\sqrt{10}}l'-\frac{1}{45}m' \right),\cr 
&(\hat{g}^{(3),\Xi_{c}}_{A})^{(\mathrm{wf})}=T_{3}
  \left(-\frac{5}{9\sqrt{15}}l'-\frac{4}{45\sqrt{6}}m'\right),\cr  
&(\hat{g}^{(3),\Omega_{c}}_{A})^{(\mathrm{wf})}= 0,
\end{align}
and the octet axial charges
\begin{align}
&(\hat{g}^{(8),\Sigma_{c}}_{A})^{(\mathrm{op})}= -\frac{1}{27\sqrt{3}}x'
 -\frac{4}{45\sqrt{3}}y',\cr
&(\hat{g}^{(8),\Xi_{c}}_{A})^{(\mathrm{op})}= \frac{1}{90\sqrt{3}}x'
 +\frac{1}{15\sqrt{3}}y', \cr 
&(\hat{g}^{(8),\Omega_{c}}_{A})^{(\mathrm{op})}= \frac{4}{45\sqrt{3}}x'
 +\frac{2}{15\sqrt{3}}y', \cr
&(\hat{g}^{(8),\Sigma_{c}}_{A})^{(\mathrm{wf})}=
  \frac{2}{3\sqrt{30}}l'-\frac{2}{15\sqrt{3}}m', \cr 
&(\hat{g}^{(8),\Xi_{c}}_{A})^{(\mathrm{wf})}=
  -\frac{1}{2\sqrt{30}}l'-\frac{2}{15\sqrt{2}}m', \cr  
&(\hat{g}^{(8),\Omega_{c}}_{A})^{(\mathrm{wf})}= -\frac{2}{15\sqrt{2}}m'.
\end{align}
Interestingly, since the baryon sextets with the spins $J'=1/2$ and
$J'=3/2$ arise from the same $N_{c}-1$ soliton with spin $J=1$, one
can easily expect that they share the same dynamics. Indeed, the
operator and wave function corrections to both the triplet and octet
axial charges for $J'=1/2$ and $J'=3/2$ are related to each other by
the overall factor $1/2$: 
\begin{align}
&(\hat{g}^{(3,8),B_{6,J'=3/2}}_{A})^{(\mathrm{wf,op})} =
  \frac{1}{2}(\hat{g}^{(3,8),B_{6,J'=1/2}}_{A})^{(\mathrm{wf,op})}. 
\end{align}

\section{Numerical results \label{sec:3}}
\begin{figure}[htp]
\includegraphics[scale=0.165]{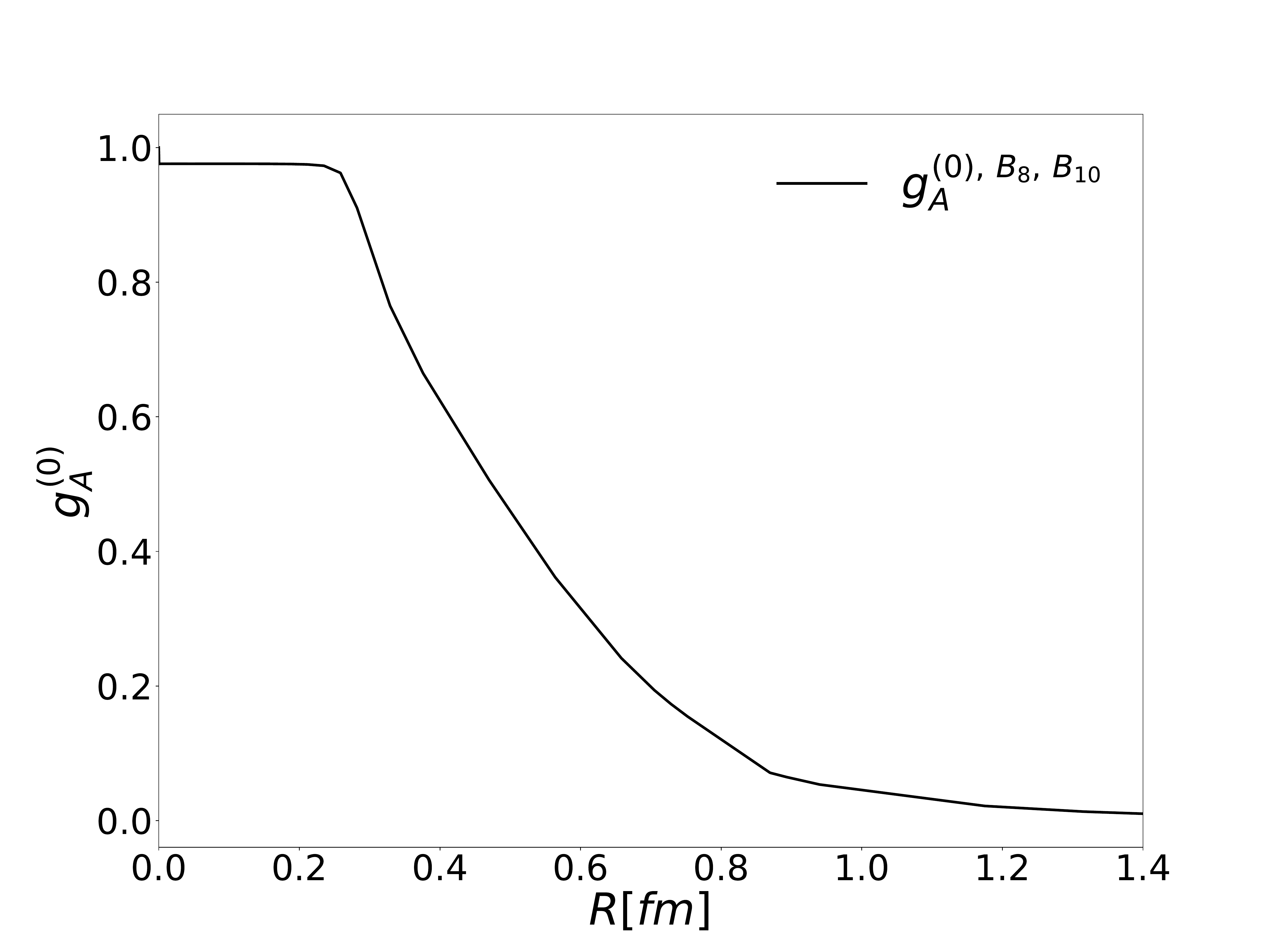}
\includegraphics[scale=0.165]{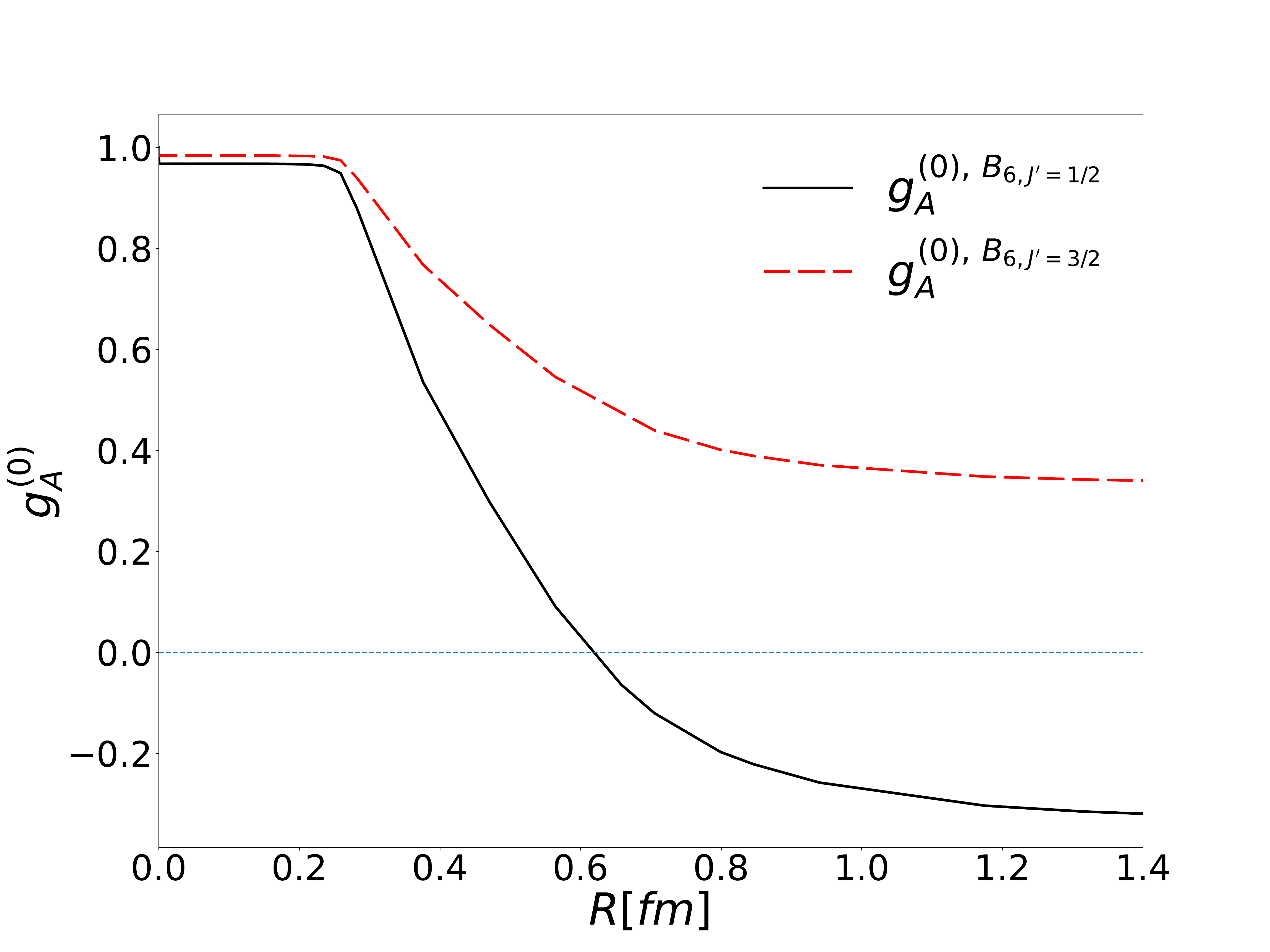}
\caption{Flavor-singlet axial charges of the light~(singly heavy)
  baryons as functions of the soliton size $R$ are depicted in the
  left~(right) panel. The profile function is taken to be a linear
  type.} 
\label{fig:1}
\end{figure}

We now present the results for the axial charges of both
the light and singly heavy baryons and discuss their physical
implications. We will first examine how the $\chi$QSM interpolates
between the naive NR quark and Skyrme
models~\cite{Praszalowicz:1995vi, Kim:1995bq}. We then provide a 
distinctive feature of the present framework in contrast to the Skyrme
model. To see those features explicitly, we draw the axial charges of
the light baryons as a function of the soliton size in the 
left panel of Fig.~\ref{fig:1}. 
Note that the flavor-singlet axial charges of the baryon octet and
baryon decuplet degenerate in the exact flavor SU(3) symmetry. When
the size of the soliton becomes zero, the value of the axial
charge in the $\chi$QSM reproduces exactly that in the naive NR quark 
model in the exact flavor SU(3) symmetry. On the other hand, the
axial charge becomes zero  
when the size of the soliton increases, which coincides with the
result from the Skyme model~\cite{Brodsky:1988ip}. Note that the
flavor-singlet axial charge has a contribution $(\mathcal{B} \
\mathrm{or} \ a_{3})$ from the imaginary part of the effective chiral
action, which does not appear in a typical Skyrme model. The
non-vanishing feature of this term in the $\chi$QSM causes the
non-zero singlet axial charge. This term is suppressed in the limit of
a large soliton size, so that the result of the Skyrme model is
reproduced. In the right panel of Fig.~\ref{fig:1}, we depict the
axial charges of the singly heavy baryons as functions of the soliton
size. Similar to the light baryon, we can restore the results from the
naive NR quark model by taking the limit of the small size
soliton. Interestingly, as the size of the soliton increases, the
axial charge of the baryon sextet with $J'=1/2$  becomes negative and
finally arrives at the value of the axial charge that solely comes
from the heavy-quark part, i.e., $\Delta{u,d,s}=0, \Delta{c}=-1/3$. On
the other hand, the axial charge of the baryon sextet with $J'=3/2$ is
always kept positive and approaches $g^{(0)}_{A}=1/3$ that solely
arises from the heavy-quark part $\Delta{u,d,s}=0, \Delta{c}=1/3$
again. These features are distinguished from the Skyrme model.    
Here we want to mention that the predictions of the $\chi$QSM lie in 
between those from the naive NR quark model and Skyrme model. 

\setlength{\tabcolsep}{1.5pt}
\renewcommand{\arraystretch}{1.5}
\begin{table}[htp]
\caption{Axial charges and their flavor decompositions of the baryon
  octet, the baryon decuplet, and the baryon sextet with $J'=1/2$,
  $J'=3/2$}  
\label{tab:1}
\begin{threeparttable}
 \begin{tabular}{ c | c c c c c c c c c c c c c} 
  \hline 
    \hline 
$J_{3}=1/2$ & $g^{(0)}_{A}$(sym)  & $g^{(0)}_{A}$&  $g^{(3)}_{A}$(sym) & $g^{(3)}_{A}$ 
& $g^{(8)}_{A}$(sym) & $g^{(8)}_{A}$ &  $\Delta{u}$(sym) & $\Delta{u}$ 
& $\Delta{d}$(sym) & $\Delta{d}$ & $\Delta{s}$(sym) & $\Delta{s}$ \\
  \hline 
$p$&$ 0.437 $ & $0.444$ & $1.163$ & $1.188$  &  $0.360$ & $0.332$ 
& $0.831$ & $0.838$ & $-0.332$ & $-0.350$ &  $ -0.062$ & $-0.044$ \\
$n$&$ 0.437$ & $0.444$ & $-1.163 $  & $-1.188$ &  $ 0.360$ & $0.332$ 
& $-0.332$ & $-0.350$  & $0.831$ & $0.838$  &  $ -0.062$ & $-0.044$ \\ 
$\Lambda$&$ 0.437$ & $0.418$ & $0.000$ & $0.000$ &  $ -0.827$ & $-0.807$ 
& $-0.093$ & $-0.094$ & $-0.093$ & $-0.094$ &  $ 0.623$ & $0.606$  \\  
$\Sigma^{+}$&$ 0.437 $ & $0.457$ & $0.893$ & $0.919$ &  $ 0.827$ & $0.794$
& $0.831$ & $0.841$ & $ -0.062$ & $-0.078$ &  $ -0.332$ & $-0.306$\\
$\Sigma^{0}$&$ 0.437  $ & $0.457$ & $0.000$ & $0.000$ &  $ 0.827$ & $0.794$
& $0.384$ & $0.381$ & $0.384$ & $0.381$ &  $ -0.332$ & $-0.306$\\
$\Sigma^{-}$&$ 0.437 $ & $0.457$ & $-0.893$ & $-0.919$ &  $ 0.827$ & $0.794$
& $-0.062$ & $-0.078$ & $0.831$ & $0.841$ &  $ -0.332$ & $-0.306$\\
$\Xi^{0}$&$ 0.437$ & $0.412$& $-0.270 $  & $-0.274$&  $ -1.187$ & $-1.173$
& $-0.332$ & $-0.338$ & $-0.062$ & $-0.064$ &  $ 0.831$ & $0.814$ \\ 
$\Xi^{-}$&$ 0.437$ & $0.412$ & $0.270$ & $0.274$ &  $ -1.187$ & $-1.173$
& $-0.062$ & $-0.064$ & $-0.332$ & $-0.338$ &  $ 0.831$ & $0.814$ \\  
\hline
$\Delta^{++}$&$ 0.437 $ & $0.461$ & $0.670 $ & $0.700$ &  $ 0.387 $ & $0.340$
& $0.592 $ & $0.602$ & $-0.077$ & $-0.098$ &  $-0.077$ & $-0.043$\\
$\Delta^{+}$&$ 0.437 $ & $0.461$ & $0.223 $ & $0.233$ &  $ 0.387 $ & $0.340$
& $0.369 $ & $0.368$ & $0.146$ & $0.135$ &  $-0.077$ & $-0.043$\\
$\Delta^{0}$&$ 0.437 $ & $0.461$ & $-0.223 $ & $-0.233$ &  $ 0.387 $ & $0.340$
& $0.146 $ & $0.135$ & $0.369$ & $0.368$ &  $-0.077$ & $-0.043$\\
$\Delta^{-}$&$ 0.437 $ & $0.461$ & $-0.670 $ & $-0.700$ &  $ 0.387 $ & $0.340$
& $-0.077 $ & $-0.098$ & $0.592$ & $0.602$ &  $-0.077$ & $-0.043$\\
$\Sigma^{*+}$&$ 0.437 $ & $0.437$ & $0.446 $ & $0.472$ &  $ 0.000 $ & $-0.021$
& $0.369$ & $0.376$ & $-0.077$ & $-0.096$ &  $0.146$ & $0.158$ \\
$\Sigma^{*0}$&$ 0.437 $ & $0.437$ & $0.000 $ & $0.000$ &  $ 0.000 $ & $-0.021$
& $0.146$ & $0.140$ & $0.146$ & $0.140$ &  $0.146$ & $0.158$ \\
$\Sigma^{*-}$&$ 0.437 $ & $0.437$ & $-0.446 $ & $-0.472$ &  $ 0.000 $ & $-0.021$
& $-0.077$ & $-0.096$ & $0.369$ & $0.376$ &  $0.146$ & $0.158$ \\
$\Xi^{*0}$&$ 0.437 $ & $0.413$ & $0.223 $ & $0.238$ &  $ -0.387 $ & $-0.390$
& $0.146$ & $0.144$ & $-0.077$ & $-0.094$ &  $0.369$ & $0.363$\\ 
$\Xi^{*-}$&$ 0.437 $ & $0.413$ & $-0.223 $ & $-0.238$ &  $ -0.387 $ & $-0.390$
& $-0.077$ & $-0.094$ & $0.146$ & $0.144$ &  $0.369$ & $0.363$\\
$\Omega^{-}$&$ 0.437 $ & $0.389$ & $0.000$ & $0.000$ &  $-0.773$ & $-0.766$
& $-0.077$ & $-0.091$ & $-0.077 $ & $-0.091$ &  $0.592$ & $0.572$ \\
\hline
 $J_{3}=1/2$ & $g^{(0)}_{A}$(sym) & $g^{(0)}_{A}$ &  $g^{(3)}_{A}$(sym) & $g^{(3)}_{A}$ 
 & $g^{(8)}_{A}$(sym) & $g^{(8)}_{A}$ & $\Delta{u}$(sym) & $\Delta{u}$ 
 & $\Delta{d}$(sym) & $\Delta{d}$ & $\Delta{s}$(sym) & $\Delta{s}$ & $\Delta{c}$ \\
  \hline 
$\Sigma^{++}_{c}$&$ 0.543 $ & $0.566$ & $1.026$ & $1.055$ &  $0.592$ & $0.568$
& $0.976$ & $0.991$ & $-0.050$ & $-0.064$ &  $-0.050$ & $-0.028$ & $-0.333$ \\  
$\Sigma^{+}_{c}$&$ 0.543 $ & $0.566$& $0.000$ & $0.000$ &  $0.592$ & $0.568$
& $0.463$ & $0.464$ & $0.463$ & $0.464$ &  $-0.050$ & $-0.028$ & $-0.333$ \\  
$\Sigma^{0}_{c}$&$ 0.543 $ & $0.566$ & $-1.026$ & $-1.055$ &  $0.592$ & $0.568$
& $-0.050$ & $-0.049$ & $0.976$ & $0.976$ &  $-0.050$ & $-0.028$ & $-0.333$ \\
$\Xi^{\prime +}_{c}$&$ 0.543 $ & $0.531$ & $0.513$ & $0.593$ &  $-0.296$ & $-0.274$
& $0.463$ & $0.505$ & $-0.050$ & $-0.087$ &  $0.463$ & $0.447$ & $-0.333$ \\  
$\Xi^{\prime 0}_{c}$&$ 0.543 $ & $0.531$ & $-0.513$ & $-0.593$ &  $-0.296$ & $-0.274$
& $-0.050$ & $-0.087$ & $0.463$ & $0.505$ &  $0.463$ & $0.447$ & $-0.333$ \\
$\Omega^{0}_{c}$&$ 0.543 $ & $0.497$ & $0.000$ & $0.000$ &  $-1.185$ & $-1.198$
& $-0.050$ & $-0.069$ & $-0.050$ & $-0.069$ &  $0.976$ & $0.968$ & $-0.333$ \\ 
\hline
$\Sigma^{* ++}_{c}$&$ 0.772 $ & $0.783$ & $0.513$ & $0.528$ &  $0.296$ & $0.284$
& $0.488$ & $0.496$ & $-0.025$ & $-0.032$ &  $-0.025$ & $-0.014$ & $0.333$ \\  
$\Sigma^{* +}_{c}$&$ 0.772 $ & $0.783$ & $0.000$ & $0.000$ &  $0.296$ & $0.284$
& $0.232$ & $0.232$ & $0.232$ & $0.232$ &  $-0.025$ & $-0.014$ & $0.333$ \\  
$\Sigma^{* 0}_{c}$&$ 0.772 $ & $0.783$ & $-0.513$ & $-0.528$ &  $0.296$ & $0.284$
& $-0.025$ & $-0.032$ & $0.488$ & $0.496$ &  $-0.025$ & $-0.014$ & $0.333$ \\ 
$\Xi^{* +}_{c}$&$ 0.772 $ & $0.766$& $0.257$ & $0.296$ &  $-0.148$ & $-0.137$
& $0.232$ & $0.253$ & $-0.025$ & $-0.044$ &  $0.232$ & $0.223$ & $0.333$ \\  
$\Xi^{* 0}_{c}$&$ 0.772 $ & $0.766$ & $-0.257$ & $-0.296$ &  $-0.148$ & $-0.137$
& $-0.025$ & $-0.044$ & $0.232$ & $0.253$ &  $0.232$ & $0.223$ & $0.333$ \\
$\Omega^{* 0}_{c}$&$ 0.772 $ & $0.748$ & $0.000$ & $0.000$ &  $-0.592$ & $-0.599$
& $-0.025$ & $-0.035$ & $-0.025$ & $-0.035$ &  $0.488$ & $0.484$ & $0.333$ \\
 \hline 
 \hline
\end{tabular} 
\end{threeparttable}   
\end{table}
It is of great interest to compare the quark spin content of both the
light and singly heavy baryons. In Table~\ref{tab:1}, we listed the 
numerical results for the axial charges and their flavor decompositions
of the baryon octet, decuplet, and sextet with $J'=1/2$ and $J'=3/2$, 
considering the exact flavor SU(3) symmetry and its breaking. It
should be noted that we choose the spin-polarization to be $J_{3}=1/2$
for the light baryons and $J'_{3}=1/2$ for the heavy baryons, so that
we can easily compare the axial charges of all the baryons.  

The values of $g^{(3),p}_{A}$ and $g^{(8),p}_{A}$ are in
agreement with those extracted from the experimental data on the
neutron $\beta$ decay and in the HSDs, respectively. Note that the
value of the $g^{(8),p}_{A}$ is still under debate. However, since
$g^{(0),p}_{A}$ cannot be obtained directly from the HSDs, one should 
relate it to the first moment of the polarized quark distribution to
extract the phenomenological value. It was obtained to be
$g^{(0),p}_{A} \sim 0.33$~\cite{Aidala:2012mv}, which is smaller than
the prediction of the $\chi$QSM $g^{(0),p}_{A} \sim 0.44$. This
discrepancy may arise from the U(1) axial anomaly, since there is no
gluonic contribution from the $\chi$QSM. We also present all the axial
charges of the baryon octet. Interestingly, the axial
charges of the $\Lambda^{0}$ and $\Sigma^{0}$ baryons have the
opposite signs to each other. Although these baryons have the same
quark content, each quark contribution shows a different value because
of the different isospin. In Table~\ref{tab:2}, we find that the
present results are consistent with those from lattice 
QCD~\cite{Alexandrou:2016xok,Lin:2007ap}, chiral perturbation 
theory~\cite{Jiang:2009sf}, the relativistic constituent quark model
(RCQM)~\cite{Choi:2010ty, Choi:2013ysa}, the perturbative chiral quark
model(PCQM)~\cite{Liu:2018jiu}.  

In Table~\ref{tab:1}, we also list the flavor-decomposed quark spin
content. The separate spin contributions of the $u$, $d$, and $s$
quarks to the nucleon are estimated empirically in
Ref.~\cite{Aidala:2012mv}: $\Delta u \sim 0.84, \Delta d \sim-0.43$,
and $\Delta s\sim -0.08$ where the normalization point is taken to be
$Q^{2}\to \infty$. Interestingly, we predict the value of $s$-quark
contribution $\Delta s \sim -0.05$ for the proton, which is comparable
to the estimation of Ref.~\cite{Aidala:2012mv},  
$\Delta s \sim -0.08$. This non-zero value of the $s$-quark
contribution breaks the Ellis-Jaffe sum rule, so that it plays an
essential role in understanding the quark spin content of the
proton. This feature can be observed for other hyperons as
well. For example, the ``pure'' sea-quark contributions to the axial 
charges of the $n,\Sigma$, and $\Xi$ are also estimated to be 
less than $-0.08$. Note that ``pure''
  sea-quark stand for the Dirac continuum in the $\chi$QSM. In
  Table~\ref{tab:2}, we find that the present results are consistent
  with those from lattice
  QCD~\cite{Liang:2018pis,Alexandrou:2017oeh,Lin:2018obj}, a global
  QCD analysis~\cite{deFlorian:2009vb}, NNPDF collaboration
data~\cite{Nocera:2014gqa}, and COMPASS data~\cite{COMPASS:2015mhb}. 

In contrast to the baryon octet, there are no experimental data on the 
axial charges of the baryon decuplet. However, since we see that
the present work has successfully described the proton axial 
charges, we can proceed to compute the axial charges of the baryon
decuplet. While the flavor-singlet axial charge of the baryon decuplet
is identical with that of the baryon octet $g^{(0),B_{10}}_{A} =
g^{(0),B_{8}}_{A}$, flavor-triplet and -octet axial charges
are different. This leads to the different combinations of the
quark-spin contributions to the baryons. Compared to the quark spin
content of the proton, the $\Delta^{+}$ isobar has a smaller value  
$\Delta u \sim 0.37$ with the same sign, whereas it has the
opposite sign and smaller value $\Delta d \sim 0.14$. The value
$\Delta s$ of $\Delta^{+}$ is almost the same as that of the 
proton. In the exact flavor SU(3) symmetry, the axial charges
of the baryon decuplet can be related to each other~\eqref{eq:C5}. In
Table~\ref{tab:3}, we compare the results for the axial charges
of the baryon decuplet with those from the other models. Here we take 
the spin polarization to be $J_{3}=3/2$ instead of $J_{3}=1/2$ to 
easily compare the present results with those from other
models. We find that the current results are consistent with those
from chiral perturbation theory~\cite{Jenkins:1991es,Jiang:2008we}, the
relativistic constituent quark model(RCQM)~\cite{Choi:2010ty,
  Choi:2013ysa}, the perturbative chiral quark
model(PCQM)~\cite{Liu:2018jiu}, and lattice
QCD~\cite{Alexandrou:2016xok}.  

In the limit of $m_{Q}\to \infty$, all the dynamics are governed by
the light quarks, whereas the heavy quark merely carries its charge
and spin. Employing the dynamical parameters obtained in the model
calculation, we can predict the axial charges of the singly heavy
baryons in the exact flavor SU(3) symmetry and with the effects of its
breaking. It should be noted that the spin of the antitriplet
($\overline{\bm{3}}$) heavy baryon is solely carried by the charm
quark, i.e., $\Delta c= 1$, as mentioned previously. The
flavor-triplet and -octet axial charges equally vanish. We want to
mention that the light quark contributions to the axial charges of the
antitriplet ($\overline{\bm{3}}$) baryons are estimated to be null in
lattice QCD~\cite{Alexandrou:2016xok},  
i.e., $g^{(0),B_{\bar{3},J'=1/2}}_{A}\sim 0.9, 
g^{(3),B_{\bar{3},J'=1/2}}_{A}\sim 0,
g^{(8),B_{\bar{3},J'=1/2}}_{A}\sim 0$, which strongly supports the
present results. 

In Table~\ref{tab:1}, we also list the axial charges of the baryon
sextet ($\bm{6}$) with $J'=1/2$ and $J'=3/2$, respectively. The 
flavor-singlet axial charges of the singly heavy baryons with $J'=1/2$
and $J'=3/2$ are respectively enhanced by $\sim 20~\%$ and $\sim 40~\%$
in comparison with those of the light baryons. In the naive NR
quark model, the charm quark yields $-1/3$ ($1/3$) for the spin  
of the baryon sextet with $J'=1/2$ ($J'=3/2$), so the light quarks
should be strongly~(weakly) polarized to keep the sum of the quark
spins unity, i.e., $g^{(0)}_{A}=1$. We expect that such a behavior 
may remain in the $\chi$QSM, though the relativistic effects further
suppress the value of $g^{(0)}_{A}$. Thus, we may conclude that the
light quarks inside the baryon sextet ($\bm{6}$) with
$J'=1/2$~($J'=3/2$) are more strongly~(weakly) polarized than 
those in the light baryons. We have encountered this 
feature already in the calculation of the heavy baryon spin out of the 
energy-momentum tensor~\cite{Kim:2020nug}. We also found an
interesting physics that the ``pure'' sea quark contributions to the
axial charges of the singly heavy baryon are smaller than
those of the light baryons. It implies that the ``pure'' sea quark
contributions inside the singly heavy baryons are relatively more 
suppressed compared to the light baryon. In addition, one can clearly
see that the light-quark contributions to the flavor-singlet  
axial charge of the baryon sextet with $J'=1/2$, i.e., $\Delta
u+\Delta d+\Delta s = 0.876$, is twice larger than  the baryon sextet
with $J'=3/2$, i.e., $\Delta u+\Delta d+\Delta s= 0.439$ in
  flavor SU(3) symmetry. Therefore, each quark contribution in both
  representations shows the same ratio except for the heavy-quark
  contributions in the exact flavor SU(3) symmetry. In
  Tables~\ref{tab:4} and~\ref{tab:5}, we compare the values of
  the axial charges of the singly heavy baryons with those from
  lattice QCD. We find that the present results are consistent with
  them~\cite{Alexandrou:2016xok}.  

\setlength{\tabcolsep}{0.5pt}
\renewcommand{\arraystretch}{1.5}
\begin{table}[htp]
\centering
\caption{Axial charges and their flavor decompositions of the baryon octet} 
\label{tab:2}
\begin{threeparttable}
\resizebox{\columnwidth}{!}
{
 \begin{tabular}{ c | c c c c c c c } 
  \hline 
    \hline 
 $J_{3}=1/2$ & $g^{(0)}_{A}$  &  $g^{(3)}_{A}$ & $g^{(8)}_{A}$ & $\Delta{u}$ & $\Delta{d}$ & $\Delta{s}$ \\
  \hline 
$p$&$ 0.444 $ & $ 1.188 $  &  $ 0.332 $ 
& $ 0.838 $ & $-0.350$  &  $ -0.044 $ \\
$N$~\cite{Jiang:2009sf}&$- $ & $ 1.18$  &  $ - $ 
& $ -$ & $ - $  &  $ -$ \\ 
$N$~\cite{Alexandrou:2017oeh}&$ - $ & $ -$  &  $ - $ 
& $ 0.415 \pm 0.013 \pm 0.002^{a} $ & $ -0.193 \pm 0.008 \pm 0.003^{a} $  &  $ -0.021 \pm 0.005 \pm 0.001^{a}$ \\
$N$~\cite{Berkowitz:2017gql}&$- $ & $ 1.278 \pm 0.021 \pm 0.026 $  &  $ - $ 
& $ -$ & $ - $  &  $ -$ \\
$N$~\cite{Liang:2018pis}&$ 0.405 \pm 0.025 \pm 0.037 $ & $ 1.254 \pm 0.016 \pm 0.030 $  &  $ 0.510 \pm 0.027 \pm 0.039 $ 
& $ 0.847 \pm 0.018 \pm 0.032 $ & $ -0.407 \pm 0.016 \pm 0.018 $  &  $ -0.035 \pm 0.006 \pm 0.007$ \\
$N$~\cite{Lin:2018obj}&$ 0.286 \pm 0.062 $ & $ 1.218 \pm 0.025 \pm 0.030$  &  $ - $ 
& $ 0.777 \pm 0.025 \pm 0.030 $ & $ -0.438 \pm 0.018 \pm 0.030 $  &  $ -0.053 \pm 0.008$ \\
$N$~\cite{deFlorian:2009vb}&$ 0.366^{+0.015}_{-0.018} $ & $ - $  &  $ - $ 
& $ 0.793^{+0.011}_{-0.012} $ & $ -0.416^{+0.011}_{-0.009} $  &  $ -0.012^{+0.020}_{-0.024}$ \\
$p$(EGBE)~\cite{Choi:2010ty}&$- $ & $ 1.15$  &  $ - $ 
& $ -$ & $ - $  &  $ -$ \\
$p$(psGBE)~\cite{Choi:2010ty}&$- $ & $ 1.15$  &  $ - $ 
& $ -$ & $ - $  &  $ -$ \\
$p$(OGE)~\cite{Choi:2010ty}&$- $ & $ 1.11$  &  $ - $ 
& $ -$ & $ - $  &  $ -$ \\
$N$~\cite{Nocera:2014gqa}&$ 0.25 \pm 0.10 $ & $ - $  &  $ - $ 
& $ 0.76 \pm 0.04 $ & $ -0.41 \pm 0.04 $  &  $ -0.10 \pm 0.08$ \\
$N$~\cite{COMPASS:2015mhb}&$[0.26,0.36] $ & $ 1.22 \pm 0.05 \pm 0.10 $  &  $ - $ 
& $ [0.82,0.85] $ & $ [-0.45,-0.42] $  &  $ [-0.11,-0.08]$ \\
$N$~\cite{Liu:2018jiu}&$- $ & $ 1.263 $  &  $ - $ 
& $ -$ & $ - $  &  $ -$ \\ 
$N$~\cite{Lin:2007ap}&$- $ & $ 1.18 \pm 0.04 \pm 0.06 $  &  $ - $ 
& $ -$ & $ - $  &  $ -$ \\
$\Lambda$&$ 0.418 $ & $0.000$  &  $ -0.807 $ 
& $-0.094$ & $-0.094$  &  $0.606$  \\  
$\Lambda$~\cite{Alexandrou:2016xok}$^{\dagger}$&$ 0.6361 \pm 0.0180 $ & $0.0851 \pm 0.0145$  &  $ -1.5169 \pm 0.238^{o} $ 
& $0.0035 \pm 0.0105 $ & $-0.0861 \pm 0.0106$  &  $ 0.7185 \pm 0.0092 $  \\  
$\Sigma^{+}$&$ 0.457  $ & $0.919$  &  $ 0.794 $ 
& $ 0.841 $ & $-0.078$  &  $-0.306$ \\
$\Sigma$~\cite{Jiang:2009sf}&$- $ & $ 0.73 $  &  $ - $ 
& $ -$ & $ - $  &  $ -$ \\
$\Sigma^{+}$(EGBE)~\cite{Choi:2010ty}&$- $ & $ 0.65^{b}$  &  $ - $ 
& $ -$ & $ - $  &  $ -$ \\
$\Sigma^{+}$(psGBE)~\cite{Choi:2010ty}&$- $ & $ 0.65^{b}$  &  $ - $ 
& $ -$ & $ - $  &  $ -$ \\
$\Sigma^{+}$(OGE)~\cite{Choi:2010ty}&$- $ & $ 0.65^{b}$  &  $ - $ 
& $ -$ & $ - $  &  $ -$ \\
$\Sigma$~\cite{Liu:2018jiu}&$- $ & $ 0.896 $  &  $ - $ 
& $ -$ & $ - $  &  $ -$ \\
$\Sigma$~\cite{Alexandrou:2016xok}$^{\dagger}$&$ 0.4984 \pm 0.0244  $ & $0.7629 \pm 0.0218$  &  $1.2885 \pm 0.0288^{o}$ 
& $ 0.7629 \pm 0.0218 $ & $ - $  &  $ -0.2634 \pm 0.0101 $ \\
$\Sigma$~\cite{Lin:2007ap}&$- $ & $ 0.450 \pm 0.021 \pm 0.027^{c} $  &  $ - $ 
& $ -$ & $ - $  &  $ -$ \\
$\Xi^{0}$&$0.412$ & $ -0.274 $  &  $ -1.173 $ 
& $ -0.338 $ & $-0.064$  &  $0.814$  \\ 
$\Xi$~\cite{Jiang:2009sf}&$- $ & $ 0.23^{d} $  &  $ - $ 
& $ -$ & $ - $  &  $ -$ \\
$\Xi^{0}$(EGBE)~\cite{Choi:2010ty}&$- $ & $ -0.21$  &  $ - $ 
& $ -$ & $ - $  &  $ -$ \\
$\Xi^{0}$(psGBE)~\cite{Choi:2010ty}&$- $ & $ -0.22$  &  $ - $ 
& $ -$ & $ - $  &  $ -$ \\
$\Xi^{0}$(OGE)~\cite{Choi:2010ty}&$- $ & $ -0.22$  &  $ - $ 
& $ -$ & $ - $  &  $ -$ \\
$\Xi$~\cite{Liu:2018jiu}&$- $ & $ -0.275 $  &  $ - $ 
& $ -$ & $ - $  &  $ -$ \\
$\Xi$~\cite{Alexandrou:2016xok}$^{\dagger}$&$ 0.6735 \pm 0.0162  $ & $-0.2479 \pm 0.0087$  &  $-2.1092 \pm 0.236^{o}$ 
& $ -0.2479 \pm 0.0087 $ & $ - $  &  $ 0.9266 \pm 0.0121 $ \\
$\Xi$~\cite{Lin:2007ap}&$- $ & $ -0.277 \pm 0.015  \pm 0.019$  &  $ - $ 
& $ -$ & $ - $  &  $ -$ \\
 \hline 
 \hline
\end{tabular}
}
\begin{tablenotes}\footnotesize
\item[a] Note that the expressions for the axial charges in 
Ref.~\cite{Alexandrou:2017oeh} are different from the present one 
by $1/2$.
\item[b] Note that the expressions for the axial charges in 
Ref.~\cite{Choi:2010ty} are different from the present one 
by $1/\sqrt{2}$.
\item[c] Note that the expressions for the axial charges in 
Ref.~\cite{Lin:2007ap} are different from the present one 
by $1/2$.
\item[d] Note that the expressions for the axial charges in 
Ref.~\cite{Jiang:2009sf} are different from the present one 
by $-1$.
\item[o] Note that the expressions for the axial charges in 
Ref.~\cite{Alexandrou:2016xok} are different from the present one 
by $\sqrt{3}$.
\item[$\dagger$] The values in Ref.~\cite{Alexandrou:2016xok} are
  obtained by averaging over various isospin partners.
\end{tablenotes}
\end{threeparttable}
\end{table}

\setlength{\tabcolsep}{5pt}
\renewcommand{\arraystretch}{1.5}
\begin{table}[htp]
\centering
\caption{Axial charges and their flavor decompositions of the baryon
  decuplet}  
\label{tab:3}
\begin{threeparttable}
 \begin{tabular}{ c | c c c c c c c } 
  \hline 
    \hline 
  $J_{3}=3/2$ & $g^{(0)}_{A}$  &  $g^{(3)}_{A}$ & $g^{(8)}_{A}$ & $\Delta{u}$ & $\Delta{d}$ & $\Delta{s}$ \\
  \hline 
$\Delta^{++}$&$1.336$ & $2.101$  &  $1.020$ 
& $ 1.790 $ & $-0.311$  &  $-0.144$ \\ 
$\Delta^{++}$(EGBE)~\cite{Choi:2010ty}&$ - $ & $ -4.48^{a} $  &  $ - $ 
& $ - $ & $ -$  &  $-$ \\
$\Delta^{++}$(psGBE)~\cite{Choi:2010ty}&$ - $ & $ -4.47^{a} $  &  $ - $ 
& $ - $ & $ -$  &  $-$ \\
$\Delta^{++}$(OGE)~\cite{Choi:2010ty}&$ - $ & $ -4.30^{a} $  &  $ - $ 
& $ - $ & $ -$  &  $-$ \\
$\Delta$~\cite{Liu:2018jiu}&$ - $ & $ 1.863 $  &  $ - $ 
& $ - $ & $ -$  &  $-$ \\
$\Delta$~\cite{Jiang:2008we}&$ - $ & $ -4.50^{a} $  &  $ - $ 
& $ - $ & $ -$  &  $-$ \\
$\Sigma^{*+}$& $1.312$ & $1.415$  &  $-0.062$ 
& $1.127$ & $-0.288$  &  $0.473$  \\
$\Sigma^{*+}$(EGBE)~\cite{Choi:2010ty}&$ - $ & $ -1.06^{b} $  &  $ - $ 
& $ - $ & $ -$  &  $-$ \\
$\Sigma^{*+}$(psGBE)~\cite{Choi:2010ty}&$ - $ & $ -1.06^{b} $  &  $ - $ 
& $ - $ & $ -$  &  $-$ \\
$\Sigma^{*+}$(OGE)~\cite{Choi:2010ty}&$ - $ & $ -1.00^{b} $  &  $ - $ 
& $ - $ & $ -$  &  $-$ \\
$\Sigma^{*}$~\cite{Liu:2018jiu}&$ - $ & $ 1.242 $  &  $ - $ 
& $-$ & $-$  &  $-$  \\ 
$\Sigma^{*}$~\cite{Alexandrou:2016xok}$^{\dagger}$&$ 1.8616 \pm 0.0498 $ & $1.1740 \pm 0.0380$  &  $-0.1925 \pm 0.0336^{o}$ 
& $1.1740 \pm 0.0380$ & $-$  &  $0.6852 \pm 0.0171$  \\ 
$\Xi^{*0}$&$1.288$ & $0.714$  &  $-1.169$ 
& $0.449$ & $-0.265$  &  $1.104$ \\  
$\Xi^{*0}$(EGBE)~\cite{Choi:2010ty}&$ - $ & $ -0.75^{c} $  &  $ - $ 
& $ - $ & $ -$  &  $-$ \\
$\Xi^{*0}$(psGBE)~\cite{Choi:2010ty}&$ - $ & $ -0.75^{c} $  &  $ - $ 
& $ - $ & $ -$  &  $-$ \\
$\Xi^{*0}$(OGE)~\cite{Choi:2010ty}&$ - $ & $ -0.70^{c} $  &  $ - $ 
& $ - $ & $ -$  &  $-$ \\
$\Xi^{*}$~\cite{Liu:2018jiu}&$ - $ & $ -0.621^{c} $  &  $ - $ 
& $-$ & $-$  &  $-$  \\
$\Xi^{*}$~\cite{Alexandrou:2016xok}$^{\dagger}$&$ 1.9571 \pm 0.0379 $ & $0.5891 \pm 0.0198$  &  $-2.1321 \pm 0.0415^{o}$ 
& $0.5891 \pm 0.0198$ & $-$  &  $1.3637 \pm 0.0245$  \\ 
$\Omega^{-}$ & $1.264$ & $0.000$  &  $-2.299$ 
& $-0.242$ & $-0.242$  &  $1.749$  \\
$\Omega^{-}$~\cite{Alexandrou:2016xok}$^{\dagger}$&$ 2.0338 \pm 0.0310 $ & $ - $  &  $-4.0677 \pm 0.0620^{o}$ 
& $-$ & $-$  &  $2.0338 \pm 0.0310$  \\ 
 \hline 
 \hline
\end{tabular} 
\begin{tablenotes}\footnotesize
\item[a] The expressions for the axial charges in
  Refs.~\cite{Jiang:2008we,Choi:2010ty} are different 
  from the present one by $-2$.
\item[b] The expressions for the axial charges in
  Ref.~\cite{Choi:2010ty} are different 
  from the present one by $-1/\sqrt{2}$.
\item[c] Note that the expressions for the axial charges in 
Ref.~\cite{Liu:2018jiu,Choi:2010ty} are different from the present one 
by $-1$.
\item[o] Note that the expressions for the axial charges in 
Ref.~\cite{Alexandrou:2016xok} are different from the present one 
by $\sqrt{3}$.
\item[$\dagger$] The values in Ref.~\cite{Alexandrou:2016xok} are
  obtained by averaging over various isospin partners.
\end{tablenotes}
\end{threeparttable}
\end{table}

\setlength{\tabcolsep}{5pt}
\renewcommand{\arraystretch}{1.5}
\begin{table}[htp]
\centering
\caption{Axial charges and their flavor decompositions of the baryon sextet with $J'=1/2$ } 
\label{tab:4}
\begin{threeparttable}
 \begin{tabular}{ c | c c c c c c c } 
  \hline 
    \hline 
 $J'_{3}=1/2$ & $g^{(0)}_{A}$  &  $g^{(3)}_{A}$ & $g^{(8)}_{A}$ & $\Delta{u}$ & $\Delta{d}$ & $\Delta{s}$ & $\Delta{c}$ \\
  \hline 
$\Sigma^{++}_{c}$ & $0.566$ & $1.055$  &  $0.568$ 
& $0.991$ & $-0.064$  &  $-0.028$  & $-0.333$ \\    
$\Sigma_{c}$~\cite{Alexandrou:2016xok}$^{\dagger}$&$ 0.4094 \pm 0.0199 $ & $0.7055 \pm 0.0191$  &  $0.7055 \pm 0.0191^{o}$ 
& $0.7055 \pm 0.0191$ & $-$  &  $-$  & $-0.2970 \pm 0.0113$ \\  
$\Xi^{\prime +}_{c}$&$0.531$ & $0.593$  &  $-0.274$ 
& $0.505$ & $-0.087$  &  $0.447$  & $-0.333$ \\  
$\Xi^{\prime}_{c}$~\cite{Alexandrou:2016xok}$^{\dagger}$&$ 0.4872 \pm 0.0127 $ & $0.3433 \pm 0.0085$  &  $-0.5596 \pm 0.0099^{o}$ 
& $0.3433 \pm 0.0085$ & $-$  &  $ 0.4539 \pm 0.0055 $  & $-0.3133 \pm 0.0069$ \\
$\Omega^{0}_{c}$& $0.497$ & $0.000$  &  $-1.198$ 
& $-0.069$ & $-0.069$  &  $0.968$  & $-0.333$ \\  
$\Omega^{0}_{c}$~\cite{Alexandrou:2016xok}$^{\dagger}$&$ 0.5428 \pm 0.0118 $ & $-$  &  $-1.7108 \pm 0.0233^{o}$ 
& $-$ & $-$  &  $ 0.8554 \pm 0.0117 $  & $-0.3125 \pm 0.0054$ \\
 \hline 
 \hline
\end{tabular}
\begin{tablenotes}\footnotesize
\item[o] Note that the expressions for the axial charges in 
Ref.~\cite{Alexandrou:2016xok} are different from the present one 
by $\sqrt{3}$.
\item[$\dagger$] The values in Ref.~\cite{Alexandrou:2016xok} are
  obtained by averaging over various isospin partners.
\end{tablenotes}
\end{threeparttable}
\end{table}

\setlength{\tabcolsep}{5pt}
\renewcommand{\arraystretch}{1.5}
\begin{table}[htp]
\centering
\caption{Axial charges and their flavor decompositions of the baryon
  sextet with $J'=3/2$}  
\label{tab:5}
\begin{threeparttable}
 \begin{tabular}{ c | c c c c c c c } 
  \hline 
    \hline 
$J'_{3}=3/2$  & $g^{(0)}_{A}$  &  $g^{(3)}_{A}$ & $g^{(8)}_{A}$ & $\Delta{u}$ & $\Delta{d}$ & $\Delta{s}$ & $\Delta{c}$ \\
  \hline 
$\Sigma^{* ++}_{c}$ & $2.349$ & $1.583$  &  $0.852$ 
& $1.487$ & $-0.096$  &  $-0.042$  & $1.000$ \\  
 $\Sigma^{*}_{c}$~\cite{Alexandrou:2016xok}$^{\dagger}$&$ 2.0004 \pm 0.0346 $ & $1.0899 \pm 0.0308$  &  $1.0899 \pm 0.0308^{o}$ 
& $1.0899 \pm 0.0308$ & $-$  &  $-$  & $0.9043 \pm 0.0090$ \\
$\Xi^{* +}_{c}$&$2.297$ & $0.889$  &  $-0.411$ 
& $0.758$ & $-0.131$  &  $0.670$  & $1.000$ \\ 
$\Xi^{*}_{c}$~\cite{Alexandrou:2016xok}$^{\dagger}$&$ 2.1192 \pm 0.0254 $ & $0.5466 \pm 0.0150$  &  $-0.7581 \pm 0.183^{o}$ 
& $0.5466 \pm 0.0150$ & $-$  &  $0.6587 \pm 0.0104$  & $0.9103 \pm 0.0075$ \\ 
$\Omega^{* 0}_{c}$&$2.245$ & $0.000$  &  $-1.797$ 
& $-0.104$ & $-0.104$  &  $1.452$  & $1.000$ \\  
$\Omega^{* 0}_{c}$~\cite{Alexandrou:2016xok}$^{\dagger}$&$ 2.1961 \pm 0.0261 $ & $-$  &  $-2.5817 \pm 0.0408^{o}$ 
& $-$ & $-$  &  $1.2904 \pm 0.0204$  & $0.9026 \pm 0.0090$ \\
 \hline 
 \hline
\end{tabular}
\begin{tablenotes}\footnotesize
\item[o] Note that the expressions for the axial charges in 
Ref.~\cite{Alexandrou:2016xok} are different from the present one 
by $\sqrt{3}$. \item[$\dagger$] The values in
Ref.~\cite{Alexandrou:2016xok} 
  are obtained by averaging over various 
isospin partners.
\end{tablenotes}
\end{threeparttable}
\end{table}

\begin{figure}[htp]
\centering
\includegraphics[scale=0.65]{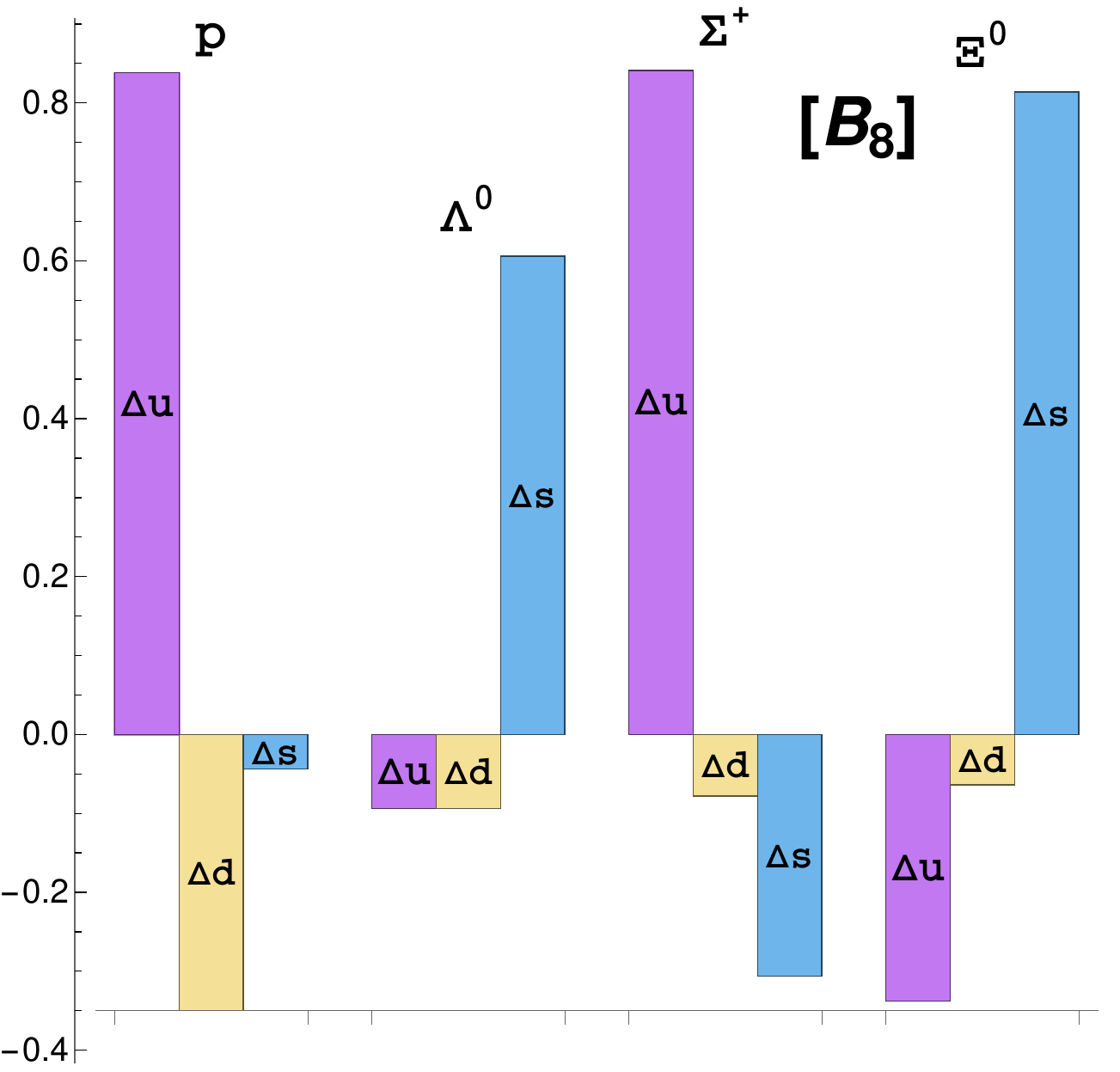}
\includegraphics[scale=0.65]{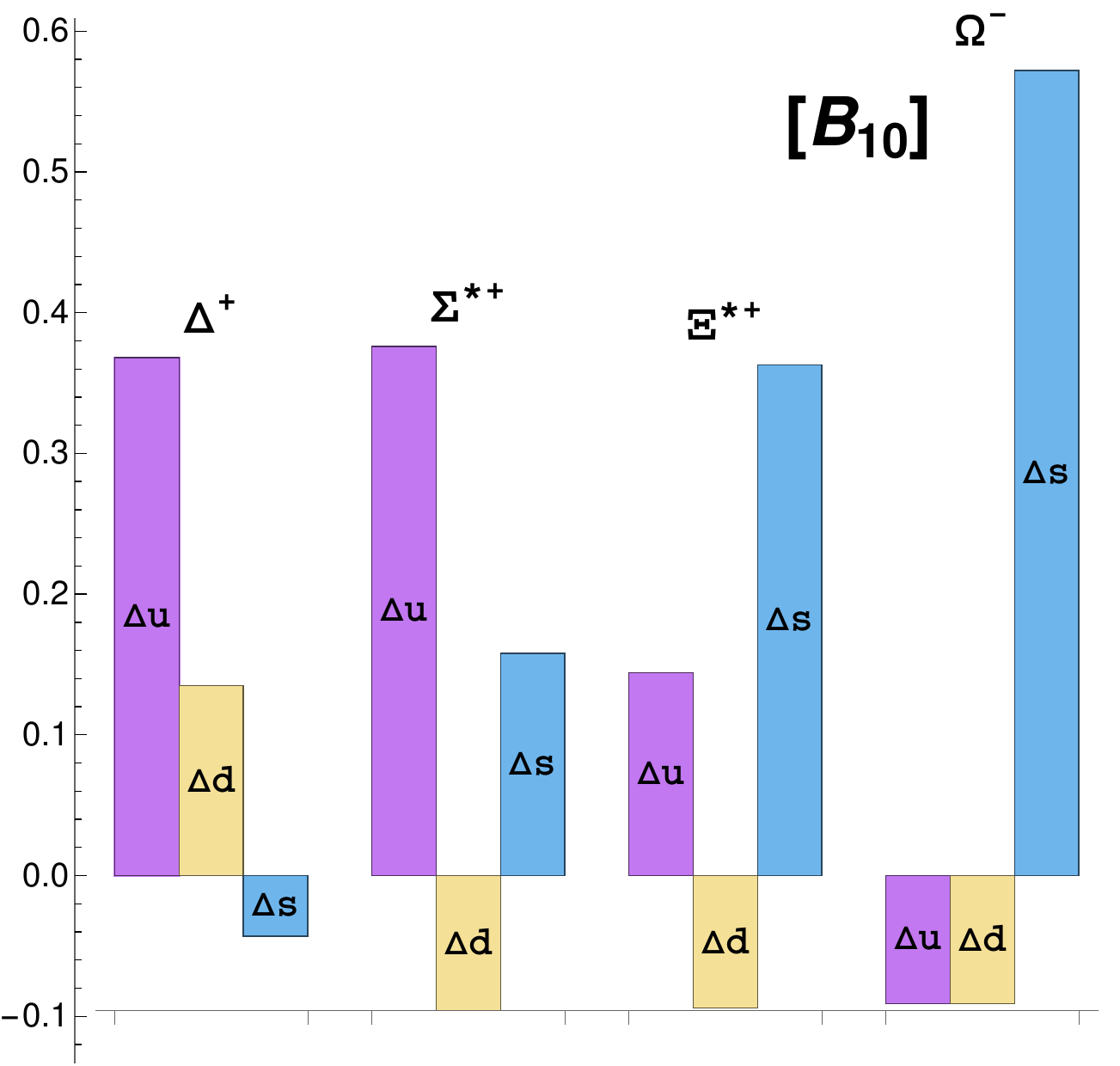}
\includegraphics[scale=0.65]{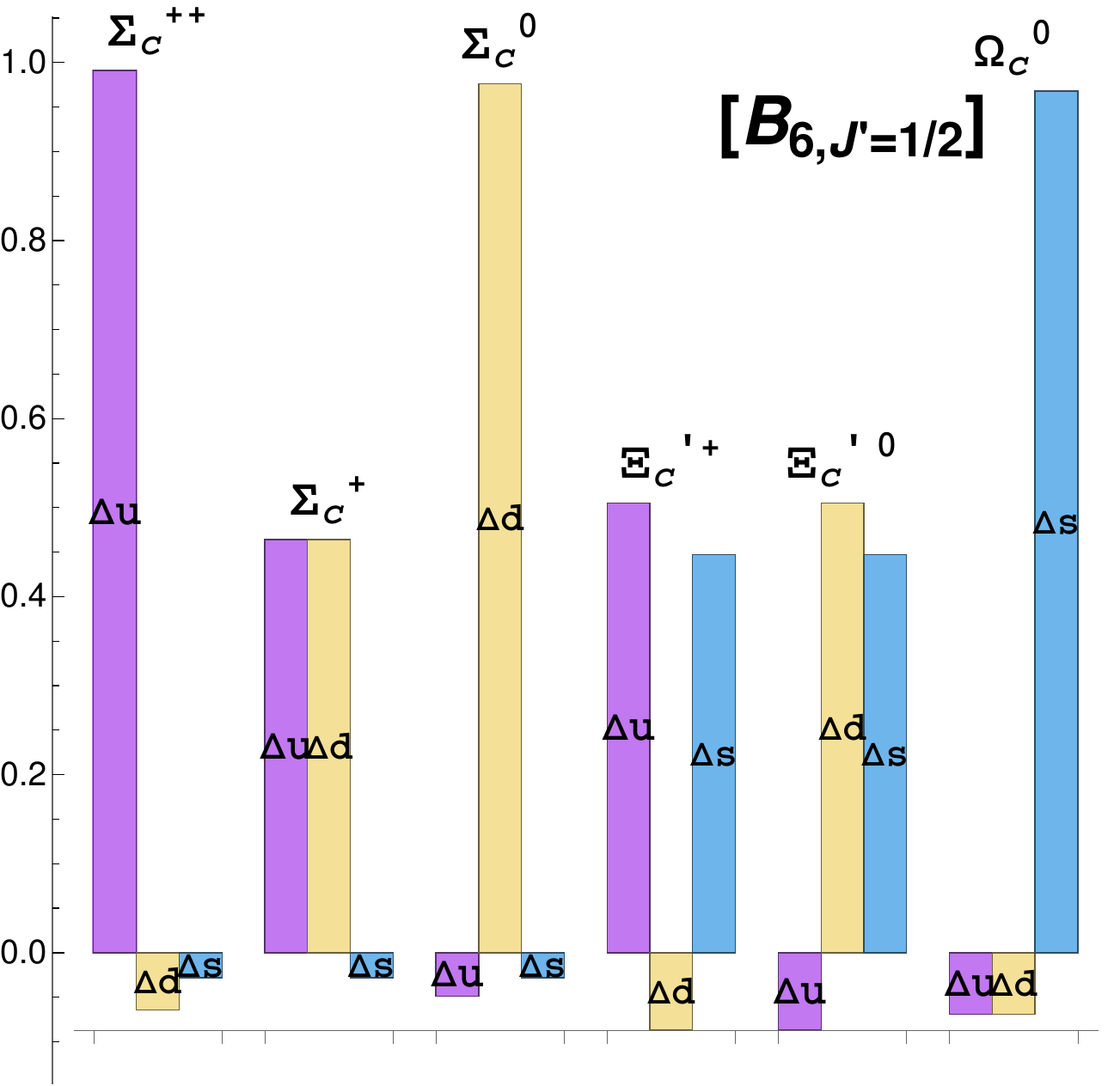}
\includegraphics[scale=0.65]{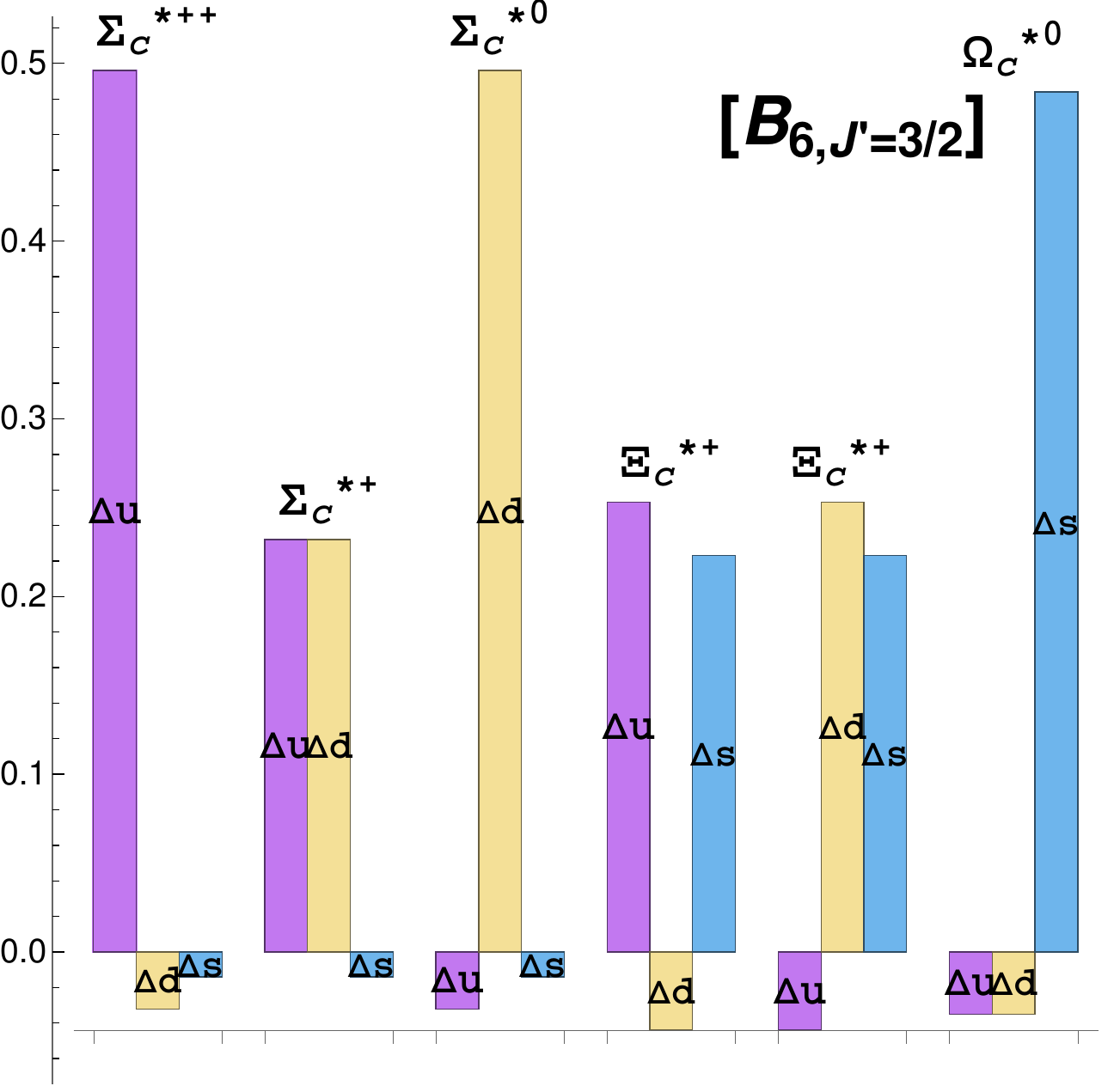}
\caption{Quark spin contents of both the lowest-lying light and singly
  heavy baryons.} 
\label{fig:2}
\end{figure}
In Fig.~\ref{fig:2}, we plot the spin content of both the 
lowest-lying light and singly heavy baryons so that one can easily
see how much fraction of the spin is carried by $u, d$, and $s$
quarks. The composite quark (say, in the naive NR quark model) spin 
content of all the baryons has the same signs. However, those of the
baryon octet have opposite signs. It means that the polarizations of
the composite quark spins inside the octet baryon are antiparallel to
each other, whereas those inside the other baryons belonging to the
other representations are parallel to each other. Meanwhile, all the
``pure'' sea quark contributions are found to be negative.  Here, one 
should keep in mind that the sea quark contributions are included in
$u,d$, and $s$ contributions. So, for the proton, $u$ and $d$ quark
contributions to its spin already include the sea quark contributions,
i.e., $\Delta\bar{u}$ and $\Delta\bar{d}$. They can not be identified
as the ``pure'' valence quark contributions. This gives a hint that
the valence quarks are more dominant in the singly heavy baryons than 
those in the light baryons. 

\section{Summary and conclusions \label{sec:4}}
In this work, we aimed at investigating the axial charges and quark
spin content of both the lowest-lying light and singly heavy baryons, 
based on the chiral quark-soliton model in the exact SU(3) symmetry
and with the effects of its breaking. We first provided the relevant
formalism for both the light and singly heavy baryons. In the case of
the light baryons, their internal dynamics are described by the pion
mean field in the presence of the $N_{c}$ valence quarks. The
dynamical parameters $a_{1\ldots6}$ related to the axial properties
were determined. On the other hand, In the limit of the infinitely
heavy quark mass ($m_{Q}\to \infty$), the heavy quark can be regarded
as a static color source, so that it merely carries its spin inside a
singly heavy baryon. Thus, the $N_c-1$ light quarks govern the
internal dynamics of the singly heavy baryon. The pion mean field was
lessened in the  presence of the $N_c-1$ valence quarks, so that
the dynamical parameters $a_{1\ldots6}$ were revised. By taking the small  
soliton-size limit, we reproduced the axial charges of both the light
and singly heavy baryons from the naive NR quark model in flavor SU(3)
symmetry. When the soliton size increased, we recovered the results
from the Skyrme model, i.e., $g^{(0)}_{A}=0$. However, those of the
heavy baryons become $g^{(0)}_{A}= \Delta c$, which is a
distinguishable feature of the chiral quark-soliton model. In fact,
the prediction of the singlet axial charges of this model lies in
between those of the naive quark and the Skyrme models. We presented
all the axial charges of both the light and singly heavy baryons and
compared them to each other. We found that the singlet axial charges
of the singly heavy baryons are enhanced in comparison with that of
the light baryon, since the heavy  quark may provide a significant
contribution to the heavy baryon spin. As a result, the light quarks
inside the heavy baryons with $J'=1/2$~($J'=3/2$) are more
strongly~(weakly) polarized compared to those inside the light
baryons. Finally, we observed that the ``pure'' sea quark 
contributions to the spin of the singly heavy baryons are always kept
to be negative and smaller than those of the light baryons. It 
implies that the sea quark contributions inside a heavy baryon are
relatively more suppressed than inside a light baryon. Thus, the 
valence quarks come into more dominant play, compared to the
light-baryon sector. We found that the present results are in good
agreement with those from the experimental and lattice QCD.

\begin{acknowledgments}
Authors are grateful to H. Y. Won for a valuable discussion. The present
work is supported by Basic Science Research Program 
through the National Research Foundation of Korea funded by the Korean
government (Ministry of Education, Science and Technology, MEST),
Grant-No. 2021R1A2C2093368, 2018R1A5A1025563, and
2019R1A2C1010443. J.-Y. Kim is supported by the Deutscher Akademischer
Austauschdienst (DAAD) doctoral scholarship.  
\end{acknowledgments}

\appendix
\section{Dynamical parameters within the $\chi$QSM \label{App:a}}
To keep using the definition of the Ref.~\cite{Jun:2020lfx}, the
dynamical parameters are expressed in terms of
$\mathcal{A},\mathcal{B},\mathcal{C}, \text{and} \ \mathcal{D}$: 
\begin{align}
& a_{1}= -\frac{1}{\sqrt{3}}\mathcal{A}
-\frac{1}{3\sqrt{2}}\frac{1}{I_1}\mathcal{D}
-\frac{2}{3\sqrt{3}}m_{s}\mathcal{H}, \quad a_{2} = -\frac{1}{\sqrt{3}}\frac{1}{I_{2}}\mathcal{C}, \quad a_{3} = \frac{1}{3}\frac{1}{I_{1}}\mathcal{B}, \cr
&a_{4} = -\frac{2}{\sqrt{3}}m_{s}\frac{K_{2}}{I_{2}}\mathcal{C}
+\frac{2}{\sqrt{3}}m_{s}\mathcal{J}, \quad
a_{5} = \frac{1}{3\sqrt{3}}m_{s}\mathcal{H}
+\frac{1}{9}\frac{K_{1}}{I_{1}}m_{s}\mathcal{B}
-\frac{1}{9}m_{s}\mathcal{I}, \cr
&a_{6} = \frac{1}{3\sqrt{3}}m_{s}\mathcal{H}
-\frac{1}{9}\frac{K_{1}}{I_{1}}m_{s}\mathcal{B}
+\frac{1}{9}m_{s}\mathcal{I},
\end{align}

with
\begin{align}
\mathcal{A} &=  
    \left[(N_{c}-N_{Q}) \phi^{\dagger}_{\mathrm{val}}
  (\bm{r}) \bm{\sigma} \cdot \bm{\tau} \phi_{\mathrm{val}}(r)
   +\; N_{c} \sum_{n} \phi^{\dagger}_{n}(\bm{r})
  \bm{\sigma} \cdot \bm{\tau}  
  \phi_{n}(\bm{r}) \mathcal{R}_{1}(E_n) \right] ,\cr
\mathcal{B} &= \left[(N_{c}-N_{Q}) \sum_{n \ne
  \mathrm{val} } \frac{1}{E_{\mathrm{val}}-E_{n}} 
  \phi^{\dagger}_{\mathrm{val}}(\bm{r}) \bm{\sigma} 
  \phi_{n}(\bm{r}) \cdot \langle n | \bm{\tau} | \mathrm{val} \rangle 
  \right. \cr
& \left. \hspace{3.9cm} 
  -\frac{1}{2}N_{c} \sum_{n,m} \phi^{\dagger}_{n}(\bm{r}) \bm{\sigma} 
  \phi_{m}(\bm{r}) \cdot \langle m | \bm{\tau} | n \rangle 
  \mathcal{R}_{5}(E_n,E_m) \right],\cr 
\mathcal{C} &= \left[(N_{c}-N_{Q}) \sum_{n_{0} \ne
  \mathrm{val} } \frac{1}{E_{\mathrm{val}}-E_{n_{0}}} 
  \phi^{\dagger}_{\mathrm{val}}(\bm{r}) \bm{\sigma} \cdot \bm{\tau} 
  \phi_{n_{0}}(\bm{r}) \langle n_{0} | \mathrm{val} \rangle \right. \cr
& \left. \hspace{3.9cm} 
  -N_{c}\sum_{n,m_{0}} \phi^{\dagger}_{n}(\bm{r}) \bm{\sigma} \cdot \bm{\tau} 
  \phi_{m_{0}}(\bm{r}) \langle m_{0} | n \rangle 
  \mathcal{R}_{5}(E_n,E_{m_{0}}) \right], \cr
\mathcal{D} &=  \left[(N_{c}-N_{Q}) \sum_{n \ne
  \mathrm{val} } \frac{\mathrm{sgn}(E_{n})}{E_{\mathrm{val}}-E_{n}} 
  \phi^{\dagger}_{\mathrm{val}}(\bm{r}) (\bm{\sigma} \times \bm{\tau})
  \phi_{n}(\bm{r}) \cdot \langle n | \bm{\tau} | \mathrm{val} 
  \rangle \right. \cr
& \left. \hspace{3.9cm} 
  + \frac{1}{2}N_{c} \sum_{n,m} \phi^{\dagger}_{n}(\bm{r}) \bm{\sigma} 
  \times \bm{\tau} \phi_{m}(\bm{r}) \cdot \langle m | \bm{\tau} 
  | n \rangle \mathcal{R}_{4}(E_n,E_m) \right],\cr
\mathcal{H} &= 
  \left[(N_{c}-N_{Q}) \sum_{n \ne
  \mathrm{val} } \frac{1}{E_{\mathrm{val}}-E_{n}} 
  \phi^{\dagger}_{\mathrm{val}}(\bm{r}) \bm{\sigma} \cdot \bm{\tau} 
  \langle \bm{r} | n \rangle \langle n | \gamma^{0}| \mathrm{val} 
  \rangle \right. \cr
& \left. \hspace{3.9cm} 
  + \frac{1}{2}N_{c} \sum_{n,m} \phi^{\dagger}_{n}(\bm{r}) \bm{\sigma} \cdot 
  \bm{\tau} \phi_{m}(\bm{r}) \langle m | \gamma^{0} | n \rangle 
  \mathcal{R}_{2}(E_n,E_m) \right], \cr
\mathcal{I} &= 
  \left[(N_{c}-N_{Q}) \sum_{n \ne
  \mathrm{val} } \frac{1}{E_{\mathrm{val}}-E_{n}} 
  \phi^{\dagger}_{\mathrm{val}}(\bm{r}) \bm{\sigma} \phi_{n}(\bm{r}) 
  \cdot \langle n | \gamma^{0} \bm{\tau} | \mathrm{val} \rangle 
  \right. \cr
& \left. \hspace{3.9cm} +\frac{1}{2}N_{c} \sum_{n,m} \phi^{\dagger}_{n}
  (\bm{r}) \bm{\sigma} \phi_{m}(\bm{r}) \cdot \langle m | \gamma^{0} 
  \bm{\tau} | n \rangle \mathcal{R}_{2}(E_n,E_m)\right],\cr
\mathcal{J} &= 
  \left[ (N_{c}-N_{Q})\sum_{n_{0} \ne
  \mathrm{val} } \frac{1}{E_{\mathrm{val}}-E_{n_{0}}} 
  \phi^{\dagger}_{\mathrm{val}}(\bm{r}) \bm{\sigma} \cdot \bm{\tau} 
  \phi_{n_{0}}(\bm{r}) \langle n_{0}| \gamma^{0} | \mathrm{val} \rangle 
  \right. \cr
& \left. \hspace{3.9cm} +N_{c} \sum_{n,m_{0}} \phi^{\dagger}_{n}(\bm{r}) 
  \bm{\sigma} \cdot \bm{\tau} \phi_{m_{0}}(\bm{r}) 
  \langle m_{0}| \gamma^{0} | n \rangle \mathcal{R}_{2}(E_n,E_{m_{0}}) 
  \right].
  \end{align}
The moments of inertia ($I_{1}, I_{2}, K_{1}, K_{2}$) are expressed
respectively as  
\begin{align}
I_{1} &= \frac{(N_{c}-N_{Q})}{6} \sum_{n\neq\mathrm{val}}
        \frac{1}{E_{n}-E_{\mathrm{val}}}\langle \mathrm{val} |
        \bm{\tau} | n \rangle \cdot \langle n | \bm{\tau} |
        \mathrm{val} \rangle + \frac{N_{c}}{12}\sum_{n,m\neq n}\langle
        m | \bm{\tau} | n \rangle \cdot \langle n | \bm{\tau} | m
        \rangle \mathcal{R}_{3}(E_n,E_m), \cr 
I_{2} &= \frac{(N_{c}-N_{Q})}{4} \sum_{n^{0}}
        \frac{1}{E_{n^{0}}-E_{\mathrm{val}}}\langle \mathrm{val} |
        n^{0} \rangle  \langle n^{0}  | \mathrm{val} \rangle
        +\frac{N_{c}}{4} \sum_{n^{0},m}\langle m | \bm{\tau} | n^{0}
        \rangle \langle n^{0}  | m \rangle
        \mathcal{R}_{3}(E_{n^{0}},E_m), \cr
K_{1} &= \frac{(N_{c}-N_{Q})}{6} \sum_{n\neq\mathrm{val}}
        \frac{1}{E_{n}-E_{\mathrm{val}}}\langle \mathrm{val} |
        \gamma^{0}\bm{\tau} | n \rangle \cdot \langle n | \bm{\tau} |
        \mathrm{val} \rangle + \frac{N_{c}}{12}\sum_{n,m\neq n}\langle
        m | \bm{\tau} | n \rangle \cdot \langle n | \gamma^{0} \bm{\tau} | m
        \rangle \mathcal{R}_{5}(E_n,E_m), \cr
K_{2} &= \frac{(N_{c}-N_{Q})}{4} \sum_{n^{0}}
        \frac{1}{E_{n^{0}}-E_{\mathrm{val}}}\langle \mathrm{val} |
        \gamma^{0}|n^{0} \rangle  \langle n^{0}  | \mathrm{val} \rangle
        +\frac{N_{c}}{4} \sum_{n^{0},m}\langle m | \bm{\tau} | n^{0}
        \rangle \langle n^{0}  | \gamma^{0}|m \rangle
        \mathcal{R}_{5}(E_{n^{0}},E_m),
\end{align}
where the regularization functions are defined by 
\begin{align}
&R_{1}(E_{n})=-\frac{1}{2\sqrt{\pi}} E_{n}\int_{1/\Lambda^{2}}^{\infty}
\frac{du}{\sqrt{u}}e^{-uE_{n}^{2}}, \cr
&\mathcal{R}_{2}(E_{n},E_{m}) = \frac{1}{2 \sqrt{\pi}} \int^{\infty}_{1/\Lambda^{2}}
   \frac{du}{\sqrt{u}} \frac{ E_{m} e^{-u E_{m}^{2}} 
  -E_{n}e^{-uE_{n}^{2}}}{E_{n} - E_{m}}, \\
&{\cal{R}}_{3}(E_{n},E_{m}) = \frac{1}{2 \sqrt{\pi}} \int^{\infty}_{1/\Lambda^{2}}
   \frac{du}{\sqrt{u}} \left[ \frac{ e^{-u E_{m}^{2}}- e^{-u
  E_{n}^{2}}}{u(E^{2}_{n} - E^{2}_{m})} -\frac{E_{m} e^{-u
  E_{m}^{2}}+E_{n} e^{-u E_{n}^{2}}}{E_{n} + E_{m}}  \right ], \cr 
&\mathcal{R}_{4}(E_{n},E_{m})=\frac{1}{2\pi}\int_{1/\Lambda^{2}}^{\infty}du\int_{0}^{1}d\alpha 
  e^{-uE_{m}^{2}+u\alpha(E_{m}^{2}-E_{n}^{2})}
  \frac{E_{m}(1-\alpha)-E_{n}\alpha}{4\pi \sqrt{\alpha(1-\alpha)}}, \cr
&{\cal{R}}_{5}(E_{n},E_{m}) =
  \frac{\mathrm{sign}(E_{n})-\mathrm{sign}(E_{m})}{2(E_{n}-E_{m})},
\end{align}
with the proper-time regulator~\cite{Christov:1995vm}. Since the
quark-loop brings about the diverging low-energy constant  in the
effective theory, we need to regularize this theory by introducing a
cutoff mass $\Lambda$. It is fixed by reproducing the experimental
value of the pion decay constant $f_{\pi}=93$ MeV. While
$|\mathrm{val}\rangle$ and $|n\rangle$ denote the state of the valence
and sea quarks with the corresponding eigenenergies $E_{\mathrm{val}}$
and $E_n$ of the single-quark Hamiltonian $h(U_c)$, respectively,
$|n_{0}\rangle$ stands for the state of sea quarks with the
corresponding eigenenergy $E_{n_{0}}$ of the free Dirac Hamiltonian
$h_{0}(1)$. The Dirac Hamiltonian is given by 
\begin{align}
h(U_{c}) = -i\gamma^{0} \gamma^{k} \partial_{k} + \gamma^{0} M
  U^{\gamma_{5}} + \gamma^{0}m, 
\label{eq:hamil}
\end{align}
where $M$ is the dynamical quark mass taken to be $420$~MeV in
this work, and the isospin symmetry breaking effects are neglected,
i.e., $m_{u}=m_{d}=m$. Note that the current quark mass $m$ is
determined by reproducing the physical pion mass
$m_{\pi}=140$~MeV. Here, the chiral field is defined as
$U^{\gamma_{5}}=\exp{i \gamma_{5} \tau^{a} \pi^{a}}$ with $U=\exp{i
  \tau^{a} \pi^{a}}$. The free Dirac Hamiltonian $h_{0}(1)$ can be
simply written by replacing $U^{\gamma_{5}}$ to unity in
Eq.~\eqref{eq:hamil}.

\section{Supplementaries}
\setlength{\tabcolsep}{0.5pt}
\renewcommand{\arraystretch}{1.5}
\begin{table}[htp]
\centering
\caption{Axial vector constant and flavor decomposition of the proton} 
\label{tab:comparep}
\begin{threeparttable}
\resizebox{\columnwidth}{!}
{
 \begin{tabular}{ c | c c c c c c c } 
\hline 
 \hline
 $J_{3}=1/2$ & $g^{(0)}_{A}$  &  $g^{(3)}_{A}$ & $g^{(8)}_{A}$ & $\Delta{u}$ & $\Delta{d}$ & $\Delta{s}$ \\
  \hline 
$p$&$ 0.444 $ & $ 1.188 $  &  $ 0.332 $ 
& $ 0.838 $ & $ -0.350 $  &  $ -0.044 $ \\ 
$N$~\cite{Jiang:2009sf}&$- $ & $ 1.18$  &  $ - $ 
& $ -$ & $ - $  &  $ -$ \\ 
$N$~\cite{Alexandrou:2017oeh}&$ - $ & $ -$  &  $ - $ 
& $ 0.415 \pm 0.013 \pm 0.002^{a} $ & $ -0.193 \pm 0.008 \pm 0.003^{a} $  &  $ -0.021 \pm 0.005 \pm 0.001^{a}$ \\
$N$~\cite{Berkowitz:2017gql}&$- $ & $ 1.278 \pm 0.021 \pm 0.026 $  &  $ - $ 
& $ -$ & $ - $  &  $ -$ \\
$N$~\cite{Liang:2018pis}&$ 0.405 \pm 0.025 \pm 0.037 $ & $ 1.254 \pm 0.016 \pm 0.030 $  &  $ 0.510 \pm 0.027 \pm 0.039 $ 
& $ 0.847 \pm 0.018 \pm 0.032 $ & $ -0.407 \pm 0.016 \pm 0.018 $  &  $ -0.035 \pm 0.006 \pm 0.007$ \\
$N$~\cite{Lin:2018obj}&$ 0.286 \pm 0.062 $ & $ 1.218 \pm 0.025 \pm 0.030$  &  $ - $ 
& $ 0.777 \pm 0.025 \pm 0.030 $ & $ -0.438 \pm 0.018 \pm 0.030 $  &  $ -0.053 \pm 0.008$ \\
$N$~\cite{deFlorian:2009vb}&$ 0.366^{+0.015}_{-0.018} $ & $ - $  &  $ - $ 
& $ 0.793^{+0.011}_{-0.012} $ & $ -0.416^{+0.011}_{-0.009} $  &  $ -0.012^{+0.020}_{-0.024}$ \\
$p$(EGBE)~\cite{Choi:2010ty}&$- $ & $ 1.15$  &  $ - $ 
& $ -$ & $ - $  &  $ -$ \\
$p$(psGBE)~\cite{Choi:2010ty}&$- $ & $ 1.15$  &  $ - $ 
& $ -$ & $ - $  &  $ -$ \\
$p$(OGE)~\cite{Choi:2010ty}&$- $ & $ 1.11$  &  $ - $ 
& $ -$ & $ - $  &  $ -$ \\
$N$~\cite{Nocera:2014gqa}&$ 0.25 \pm 0.10 $ & $ - $  &  $ - $ 
& $ 0.76 \pm 0.04 $ & $ -0.41 \pm 0.04 $  &  $ -0.10 \pm 0.08$ \\
$N$~\cite{COMPASS:2015mhb}&$[0.26,0.36] $ & $ 1.22 \pm 0.05 \pm 0.10 $  &  $ - $ 
& $ [0.82,0.85] $ & $ [-0.45,-0.42] $  &  $ [-0.11,-0.08]$ \\
$N$~\cite{Liu:2018jiu}&$- $ & $ 1.263 $  &  $ - $ 
& $ -$ & $ - $  &  $ -$ \\ 
$N$~\cite{Lin:2007ap}&$- $ & $ 1.18 \pm 0.04 \pm 0.06 $  &  $ - $ 
& $ -$ & $ - $  &  $ -$ \\
\hline 
 \hline
 \end{tabular} 
}
\begin{tablenotes}\footnotesize
\item[a] Note that the expressions for the axial charges in 
Ref.~\cite{Alexandrou:2017oeh} are different from the present one 
by $1/2$.
\end{tablenotes}
\end{threeparttable}   
\end{table}

\setlength{\tabcolsep}{5.0pt}
\renewcommand{\arraystretch}{1.5}
\begin{table}[htp]
\centering
\caption{Axial vector constant and flavor decomposition of $\Lambda$} 
\label{tab:compareL}
\begin{threeparttable}
\resizebox{\columnwidth}{!}
{
 \begin{tabular}{ c | c c c c c c c } 
\hline 
 \hline
 $J_{3}=1/2$ & $g^{(0)}_{A}$  &  $g^{(3)}_{A}$ & $g^{(8)}_{A}$ & $\Delta{u}$ & $\Delta{d}$ & $\Delta{s}$ \\
  \hline 
  $\Lambda$&$ 0.418 $ & $0.000$  &  $ -0.807 $ 
& $-0.094$ & $-0.094$  &  $ 0.606 $  \\  
$\Lambda$~\cite{Alexandrou:2016xok}$^{\dagger}$& $0.6361 \pm 0.0180$ & $0.0851 \pm 0.0145$ 
&  $ -1.5169 \pm 0.238^{o} $ 
& $0.0035 \pm 0.0105 $ & $-0.0861 \pm 0.0106$  &  $ 0.7185 \pm 0.0092 $  \\  
  \hline 
 \hline
 \end{tabular} 
}
\begin{tablenotes}\footnotesize
\item[o] Note that the expressions for the axial charges in 
Ref.~\cite{Alexandrou:2016xok} are different from the present one 
by $\sqrt{3}$.
\item[$\dagger$] The values in Ref.~\cite{Alexandrou:2016xok} are
  obtained by averaging over various isospin partners.
\end{tablenotes}
\end{threeparttable}   
\end{table}

\setlength{\tabcolsep}{7.0pt}
\renewcommand{\arraystretch}{1.5}
\begin{table}[htp]
\centering
\caption{Axial vector constant and flavor decomposition of $\Sigma^{+}$} 
\label{tab:compareS}
\begin{threeparttable}
\resizebox{\columnwidth}{!}
{
 \begin{tabular}{ c | c c c c c c c } 
\hline 
 \hline
 $J_{3}=1/2$ & $g^{(0)}_{A}$  &  $g^{(3)}_{A}$ & $g^{(8)}_{A}$ & $\Delta{u}$ & $\Delta{d}$ & $\Delta{s}$ \\
  \hline 
$\Sigma^{+}$&$ 0.457  $ & $ 0.919 $  &  $ 0.794 $ 
& $ 0.841 $ & $ -0.078 $  &  $ -0.306 $ \\
$\Sigma$~\cite{Jiang:2009sf}&$- $ & $ 0.73 $  &  $ - $ 
& $ -$ & $ - $  &  $ -$ \\
$\Sigma^{+}$(EGBE)~\cite{Choi:2010ty}&$- $ & $ 0.65^{b}$  &  $ - $ 
& $ -$ & $ - $  &  $ -$ \\
$\Sigma^{+}$(psGBE)~\cite{Choi:2010ty}&$- $ & $ 0.65^{b}$  &  $ - $ 
& $ -$ & $ - $  &  $ -$ \\
$\Sigma^{+}$(OGE)~\cite{Choi:2010ty}&$- $ & $ 0.65^{b}$  &  $ - $ 
& $ -$ & $ - $  &  $ -$ \\
$\Sigma$~\cite{Liu:2018jiu}&$- $ & $ 0.896 $  &  $ - $ 
& $ -$ & $ - $  &  $ -$ \\
$\Sigma$~\cite{Alexandrou:2016xok}$^{\dagger}$&$ 0.4984 \pm 0.0244  $ & $0.7629 \pm 0.0218$  &  $1.2885 \pm 0.0288^{o}$ 
& $ 0.7629 \pm 0.0218 $ & $ - $  &  $ -0.2634 \pm 0.0101 $ \\
$\Sigma$~\cite{Lin:2007ap}&$- $ & $ 0.450 \pm 0.021 \pm 0.027^{c} $  &  $ - $ 
& $ -$ & $ - $  &  $ -$ \\
  \hline 
 \hline
 \end{tabular} 
}
\begin{tablenotes}\footnotesize
\item[b] Note that the expressions for the axial charges in 
Ref.~\cite{Choi:2010ty} are different from the present one 
by $1/\sqrt{2}$.
\item[c] Note that the expressions for the axial charges in 
Ref.~\cite{Lin:2007ap} are different from the present one 
by $1/2$.
\item[o] Note that the expressions for the axial charges in 
Ref.~\cite{Alexandrou:2016xok} are different from the present one 
by $\sqrt{3}$.
\item[$\dagger$] The values in Ref.~\cite{Alexandrou:2016xok} are
  obtained by averaging over various 
isospin partners.
\end{tablenotes}
\end{threeparttable}   
\end{table}

\setlength{\tabcolsep}{8.0pt}
\renewcommand{\arraystretch}{1.5}
\begin{table}[htp]
\centering
\caption{Axial vector constant and flavor decomposition of $\Xi^{0}$} 
\label{tab:compareX}
\begin{threeparttable}
\resizebox{\columnwidth}{!}
{
 \begin{tabular}{ c | c c c c c c c } 
\hline 
 \hline
 $J_{3}=1/2$ & $g^{(0)}_{A}$  &  $g^{(3)}_{A}$ & $g^{(8)}_{A}$ & $\Delta{u}$ & $\Delta{d}$ & $\Delta{s}$ \\
  \hline 
$\Xi^{0}$&$ 0.412 $ & $ -0.274 $  &  $ -1.173 $ 
& $ -0.338 $ & $ -0.064 $  &  $ 0.814 $  \\ 
$\Xi$~\cite{Jiang:2009sf}&$- $ & $ 0.23^{c} $  &  $ - $ 
& $ -$ & $ - $  &  $ -$ \\
$\Xi^{0}$(EGBE)~\cite{Choi:2010ty}&$- $ & $ -0.21$  &  $ - $ 
& $ -$ & $ - $  &  $ -$ \\
$\Xi^{0}$(psGBE)~\cite{Choi:2010ty}&$- $ & $ -0.22$  &  $ - $ 
& $ -$ & $ - $  &  $ -$ \\
$\Xi^{0}$(OGE)~\cite{Choi:2010ty}&$- $ & $ -0.22$  &  $ - $ 
& $ -$ & $ - $  &  $ -$ \\
$\Xi$~\cite{Liu:2018jiu}&$- $ & $ -0.275 $  &  $ - $ 
& $ -$ & $ - $  &  $ -$ \\
$\Xi$~\cite{Alexandrou:2016xok}$^{\dagger}$&$ 0.6735 \pm 0.0162  $ & $-0.2479 \pm 0.0087$  &  $-2.1092 \pm 0.236^{o}$ 
& $ -0.2479 \pm 0.0087 $ & $ - $  &  $ 0.9266 \pm 0.0121 $ \\
$\Xi$~\cite{Lin:2007ap}&$- $ & $ -0.277 \pm 0.015  \pm 0.019$  &  $ - $ 
& $ -$ & $ - $  &  $ -$ \\
  \hline 
 \hline
 \end{tabular} 
}
\begin{tablenotes}\footnotesize
\item[c] Note that the expressions for the axial charges in 
Ref.~\cite{Jiang:2009sf} are different from the present one 
by $-1$.
\item[o] Note that the expressions for the axial charges in 
Ref.~\cite{Alexandrou:2016xok} are different from the present one 
by $\sqrt{3}$.
\item[$\dagger$] The values in Ref.~\cite{Alexandrou:2016xok} are
  obtained by averaging over various isospin partners.
\end{tablenotes}
\end{threeparttable}   
\end{table}

\setlength{\tabcolsep}{14.0pt}
\renewcommand{\arraystretch}{1.5}
\begin{table}[htp]
\centering
\caption{Axial vector constant and flavor decomposition of $\Delta^{++}$} 
\label{tab:compareD}
\begin{threeparttable}
\resizebox{\columnwidth}{!}
{
 \begin{tabular}{ c | c c c c c c c } 
\hline 
 \hline
 $J_{3}=3/2$ & $g^{(0)}_{A}$  &  $g^{(3)}_{A}$ & $g^{(8)}_{A}$ & $\Delta{u}$ & $\Delta{d}$ & $\Delta{s}$ \\
  \hline 
$\Delta^{++}$& $1.336$ & $ 2.101 $  &  $ 1.020 $ 
& $ 1.790 $ & $-0.311$  &  $-0.144$ \\
$\Delta^{++}$(EGBE)~\cite{Choi:2010ty}&$ - $ & $ -4.48^{*} $  &  $ - $ 
& $ - $ & $ -$  &  $-$ \\
$\Delta^{++}$(psGBE)~\cite{Choi:2010ty}&$ - $ & $ -4.47^{*} $  &  $ - $ 
& $ - $ & $ -$  &  $-$ \\
$\Delta^{++}$(OGE)~\cite{Choi:2010ty}&$ - $ & $ -4.30^{*} $  &  $ - $ 
& $ - $ & $ -$  &  $-$ \\
$\Delta$~\cite{Liu:2018jiu}&$ - $ & $ 1.863 $  &  $ - $ 
& $ - $ & $ -$  &  $-$ \\
$\Delta$~\cite{Jiang:2008we}&$ - $ & $ -4.50^{*} $  &  $ - $ 
& $ - $ & $ -$  &  $-$ \\
    \hline 
 \hline
 \end{tabular} 
}
\begin{tablenotes}\footnotesize
\item[*] The expressions for the axial charges in
  Refs.~\cite{Jiang:2008we,Choi:2010ty} are different 
  from the present one by $-2$.
\end{tablenotes}
\end{threeparttable}   
\end{table}

\setlength{\tabcolsep}{1.0pt}
\renewcommand{\arraystretch}{1.5}
\begin{table}[htp]
\centering
\caption{Axial vector constant and flavor decomposition of $\Sigma^{*+}$} 
\label{tab:compareSs}
\begin{threeparttable}
\resizebox{\columnwidth}{!}
{
 \begin{tabular}{ c | c c c c c c c } 
\hline 
 \hline
 $J_{3}=3/2$ & $g^{(0)}_{A}$  &  $g^{(3)}_{A}$ & $g^{(8)}_{A}$ & $\Delta{u}$ & $\Delta{d}$ & $\Delta{s}$ \\
  \hline 
$\Sigma^{*+}$&$ 1.312 $ & $1.415$  &  $-0.062$ 
& $1.127$ & $-0.288$  &  $0.473$  \\
  $\Sigma^{*+}$(EGBE)~\cite{Choi:2010ty}&$ - $ & $ -1.06^{dd} $  &  $ - $ 
& $ - $ & $ -$  &  $-$ \\
$\Sigma^{*+}$(psGBE)~\cite{Choi:2010ty}&$ - $ & $ -1.06^{dd} $  &  $ - $ 
& $ - $ & $ -$  &  $-$ \\
$\Sigma^{*+}$(OGE)~\cite{Choi:2010ty}&$ - $ & $ -1.00^{dd} $  &  $ - $ 
& $ - $ & $ -$  &  $-$ \\
$\Sigma^{*}$~\cite{Liu:2018jiu}&$ - $ & $ 1.242 $  &  $ - $ 
& $-$ & $-$  &  $-$  \\ 
$\Sigma^{*}$~\cite{Alexandrou:2016xok}$^{\dagger}$&$ 1.8616 \pm 0.0498 $ & $1.1740 \pm 0.0380$  &  $-0.1925 \pm 0.0336^{o}$ 
& $1.1740 \pm 0.0380$ & $-$  &  $0.6852 \pm 0.0171$  \\
    \hline 
 \hline
 \end{tabular} 
}
\begin{tablenotes}\footnotesize
\item[dd] The expressions for the axial charges in
  Ref.~\cite{Choi:2010ty} are different 
  from the present one by $-1/\sqrt{2}$.
\item[o] Note that the expressions for the axial charges in 
Ref.~\cite{Alexandrou:2016xok} are different from the present one 
by $\sqrt{3}$.
\item[$\dagger$] The values in Ref.~\cite{Alexandrou:2016xok} are obtained by averaging over various
isospin partners.
\end{tablenotes}
\end{threeparttable}   
\end{table}
\setlength{\tabcolsep}{0.5pt}
\renewcommand{\arraystretch}{1.5}
\begin{table}[htp]
\centering
\caption{Axial vector constant and flavor decomposition of $\Xi^{*0}$} 
\label{tab:compareXis}
\begin{threeparttable}
\resizebox{\columnwidth}{!}
{
 \begin{tabular}{ c | c c c c c c c } 
\hline 
 \hline
 $J_{3}=3/2$ & $g^{(0)}_{A}$  &  $g^{(3)}_{A}$ & $g^{(8)}_{A}$ & $\Delta{u}$ & $\Delta{d}$ & $\Delta{s}$ \\
  \hline 
$\Xi^{*0}$&$ 1.288 $ & $0.714$  &  $-1.169$ 
& $0.449$ & $-0.265$  &  $1.104$ \\
$\Xi^{*0}$(EGBE)~\cite{Choi:2010ty}&$ - $ & $ -0.75^{dc} $  &  $ - $ 
& $ - $ & $ -$  &  $-$ \\
$\Xi^{*0}$(psGBE)~\cite{Choi:2010ty}&$ - $ & $ -0.75^{dc} $  &  $ - $ 
& $ - $ & $ -$  &  $-$ \\
$\Xi^{*0}$(OGE)~\cite{Choi:2010ty}&$ - $ & $ -0.70^{dc} $  &  $ - $ 
& $ - $ & $ -$  &  $-$ \\
$\Xi^{*}$~\cite{Liu:2018jiu}&$ - $ & $ -0.621^{dc} $  &  $ - $ 
& $-$ & $-$  &  $-$  \\
$\Xi^{*}$~\cite{Alexandrou:2016xok}$^{\dagger}$&$ 1.9571 \pm 0.0379 $ & $0.5891 \pm 0.0198$  &  $-2.1321 \pm 0.0415^{o}$ 
& $0.5891 \pm 0.0198$ & $-$  &  $1.3637 \pm 0.0245$  \\ 
    \hline 
 \hline
 \end{tabular} 
}
\begin{tablenotes}\footnotesize
\item[dc] Note that the expressions for the axial charges in 
Refs.~\cite{Liu:2018jiu,Choi:2010ty} are different from the present one 
by $-1$.
\item[o] Note that the expressions for the axial charges in 
Ref.~\cite{Alexandrou:2016xok} are different from the present one 
by $\sqrt{3}$.
\item[$\dagger$] The values in Ref.~\cite{Alexandrou:2016xok} are
  obtained by averaging over various isospin partners.
\end{tablenotes}
\end{threeparttable}   
\end{table}
\setlength{\tabcolsep}{12pt}
\renewcommand{\arraystretch}{1.5}
\begin{table}[htp]
\centering
\caption{Axial vector constant and flavor decomposition of $\Omega^{-}$} 
\label{tab:compareOmega}
\begin{threeparttable}
\resizebox{\columnwidth}{!}
{
 \begin{tabular}{ c | c c c c c c c } 
\hline 
 \hline
 $J_{3}=3/2$ & $g^{(0)}_{A}$  &  $g^{(3)}_{A}$ & $g^{(8)}_{A}$ & $\Delta{u}$ & $\Delta{d}$ & $\Delta{s}$ \\
  \hline 
$\Omega^{-}$&$ 1.264 $ & $0.000$  &  $-2.299$ 
& $-0.242$ & $-0.242$  &  $1.749$  \\
$\Omega^{-}$~\cite{Alexandrou:2016xok}$^{\dagger}$&$ 2.0338 \pm 0.0310 $ & $ - $  &  $-4.0677 \pm 0.0620^{o}$ 
& $-$ & $-$  &  $2.0338 \pm 0.0310$  \\ 
    \hline 
 \hline
 \end{tabular} 
}
\begin{tablenotes}\footnotesize
\item[o] Note that the expressions for the axial charges in 
Ref.~\cite{Alexandrou:2016xok} are different from the present one 
by $\sqrt{3}$.
\item[$\dagger$] The values in Ref.~\cite{Alexandrou:2016xok} are
  obtained by averaging over various isospin partners.
\end{tablenotes}
\end{threeparttable}   
\end{table}

\setlength{\tabcolsep}{1.0pt}
\renewcommand{\arraystretch}{1.5}
\begin{table}[htp]
\centering
\caption{Axial vector constant and flavor decomposition of $\Sigma^{++}_{c}$} 
\label{tab:compareoctSigmac}
\begin{threeparttable}
\resizebox{\columnwidth}{!}
{
 \begin{tabular}{ c | c c c c c c c c} 
\hline 
 \hline
 $J_{3}'=1/2$ & $g^{(0)}_{A}$  &  $g^{(3)}_{A}$ & $g^{(8)}_{A}$ & $\Delta{u}$ & $\Delta{d}$ & $\Delta{s}$ & $\Delta{c}$ \\
  \hline 
$\Sigma^{++}_{c}$&$ 0.566 $ & $1.055$  &  $0.568$ 
& $0.991$ & $-0.064$  &  $-0.028$  & $-0.333$ \\
$\Sigma_{c}$~\cite{Alexandrou:2016xok}$^{\dagger}$&$ 0.4094 \pm 0.0199 $ & $0.7055 \pm 0.0191$  &  $0.7055 \pm 0.0191^{o}$ 
& $0.7055 \pm 0.0191$ & $-$  &  $-$  & $-0.2970 \pm 0.0113$ \\
    \hline 
 \hline
 \end{tabular} 
}
\begin{tablenotes}\footnotesize
\item[o] Note that the expressions for the axial charges in 
Ref.~\cite{Alexandrou:2016xok} are different from the present one 
by $\sqrt{3}$.
\item[$\dagger$] The values in Ref.~\cite{Alexandrou:2016xok} are
  obtained by averaging over various isospin partners.
\end{tablenotes}
\end{threeparttable}   
\end{table}
\setlength{\tabcolsep}{0.5pt}
\renewcommand{\arraystretch}{1.5}
\begin{table}[htp]
\centering
\caption{Axial vector constant and flavor decomposition of $\Xi^{\prime +}_{c}$} 
\label{tab:compareoctXic}
\begin{threeparttable}
\resizebox{\columnwidth}{!}
{
 \begin{tabular}{ c | c c c c c c c c} 
\hline 
 \hline
 $J_{3}'=1/2$ & $g^{(0)}_{A}$  &  $g^{(3)}_{A}$ & $g^{(8)}_{A}$ & $\Delta{u}$ & $\Delta{d}$ & $\Delta{s}$ & $\Delta{c}$ \\
  \hline 
$\Xi^{\prime +}_{c}$&$ 0.531 $ & $0.593$  &  $-0.274$ 
& $0.505$ & $-0.087$  &  $0.447$  & $-0.333$ \\
$\Xi^{\prime}_{c}$~\cite{Alexandrou:2016xok}$^{\dagger}$&$ 0.4872 \pm 0.0127 $ & $0.3433 \pm 0.0085$  &  $-0.5596 \pm 0.0099^{o}$ 
& $0.3433 \pm 0.0085$ & $-$  &  $ 0.4539 \pm 0.0055 $  & $-0.3133 \pm 0.0069$ \\
    \hline 
 \hline
 \end{tabular} 
}
\begin{tablenotes}\footnotesize
\item[o] Note that the expressions for the axial charges in 
Ref.~\cite{Alexandrou:2016xok} are different from the present one 
by $\sqrt{3}$.
\item[$\dagger$] The values in Ref.~\cite{Alexandrou:2016xok} are obtained by averaging over various
isospin partners.
\end{tablenotes}\end{threeparttable}   
\end{table}
\setlength{\tabcolsep}{7.0pt}
\renewcommand{\arraystretch}{1.5}
\begin{table}[htp]
\centering
\caption{Axial vector constant and flavor decomposition of $\Omega^{0}_{c}$} 
\label{tab:compareOmegac}
\begin{threeparttable}
\resizebox{\columnwidth}{!}
{
 \begin{tabular}{ c | c c c c c c c c} 
\hline 
 \hline
 $J_{3}'=1/2$ & $g^{(0)}_{A}$  &  $g^{(3)}_{A}$ & $g^{(8)}_{A}$ & $\Delta{u}$ & $\Delta{d}$ & $\Delta{s}$ & $\Delta{c}$ \\
  \hline 
$\Omega^{0}_{c}$&$ 0.497 $ & $0.00$  &  $-1.198$ 
& $-0.069$ & $-0.069$  &  $0.968$  & $-0.333$ \\  
$\Omega^{0}_{c}$~\cite{Alexandrou:2016xok}$^{\dagger}$&$ 0.5428 \pm 0.0118 $ & $-$  &  $-1.7108 \pm 0.0233^{o}$ 
& $-$ & $-$  &  $ 0.8554 \pm 0.0117 $  & $-0.3125 \pm 0.0054$ \\
    \hline 
 \hline
 \end{tabular} 
}
\begin{tablenotes}\footnotesize
\item[o] Note that the expressions for the axial charges in 
Ref.~\cite{Alexandrou:2016xok} are different from the present one 
by $\sqrt{3}$.
\item[$\dagger$] The values in Ref.~\cite{Alexandrou:2016xok} are
  obtained by averaging over various isospin partners.
\end{tablenotes}\end{threeparttable}   
\end{table}

\setlength{\tabcolsep}{1.5pt}
\renewcommand{\arraystretch}{1.5}
\begin{table}[htp]
\centering
\caption{Axial vector constant and flavor decomposition of $\Sigma^{* ++}_{c}$} 
\label{tab:compareSigmacs}
\begin{threeparttable}
\resizebox{\columnwidth}{!}
{
 \begin{tabular}{ c | c c c c c c c c} 
\hline 
 \hline
 $J_{3}'=3/2$ & $g^{(0)}_{A}$  &  $g^{(3)}_{A}$ & $g^{(8)}_{A}$ & $\Delta{u}$ & $\Delta{d}$ & $\Delta{s}$ & $\Delta{c}$ \\
  \hline 
$\Sigma^{* ++}_{c}$&$ 2.349 $ & $1.583$  &  $0.852$ 
& $1.487$ & $-0.096$  &  $-0.042$  & $1.000$ \\  
$\Sigma^{*}_{c}$~\cite{Alexandrou:2016xok}$^{\dagger}$&$ 2.0004 \pm 0.0346 $ & $1.0899 \pm 0.0308$  &  $1.0899 \pm 0.0308^{o}$ 
& $1.0899 \pm 0.0308$ & $-$  &  $-$  & $0.9043 \pm 0.0090$ \\
    \hline 
 \hline
 \end{tabular} 
}
\begin{tablenotes}\footnotesize
\item[o] Note that the expressions for the axial charges in 
Ref.~\cite{Alexandrou:2016xok} are different from the present one 
by $\sqrt{3}$.
\item[$\dagger$] The values in Ref.~\cite{Alexandrou:2016xok} are obtained by averaging over various
isospin partners.
\end{tablenotes}\end{threeparttable}   
\end{table}
\setlength{\tabcolsep}{1.0pt}
\renewcommand{\arraystretch}{1.5}
\begin{table}[htp]
\centering
\caption{Axial vector constant and flavor decomposition of $\Xi^{* +}_{c}$} 
\label{tab:compareXics}
\begin{threeparttable}
\resizebox{\columnwidth}{!}
{
 \begin{tabular}{ c | c c c c c c c c} 
\hline 
 \hline
 $J_{3}'=3/2$ & $g^{(0)}_{A}$  &  $g^{(3)}_{A}$ & $g^{(8)}_{A}$ &
 $\Delta{u}$ & $\Delta{d}$ & $\Delta{s}$ & $\Delta{c}$ \\ 
  \hline 
$\Xi^{* +}_{c}$&$ 2.297 $ & $0.889$  &  $-0.411$ 
& $0.758$ & $-0.131$  &  $0.670$  & $1.000$ \\
$\Xi^{*}_{c}$~\cite{Alexandrou:2016xok}$^{\dagger}$&$ 2.1192 \pm 0.0254 $ & $0.5466 \pm 0.0150$  &  $-0.7581 \pm 0.183^{o}$ 
& $0.5466 \pm 0.0150$ & $-$  &  $0.6587 \pm 0.0104$  & $0.9103 \pm 0.0075$ \\
    \hline 
 \hline
 \end{tabular} 
}
\begin{tablenotes}\footnotesize
\item[o] Note that the expressions for the axial charges in 
Ref.~\cite{Alexandrou:2016xok} are different from the present one 
by $\sqrt{3}$.
\item[$\dagger$] The values in Ref.~\cite{Alexandrou:2016xok} are obtained by averaging over various
isospin partners.
\end{tablenotes}
\end{threeparttable}   
\end{table}
\setlength{\tabcolsep}{7.0pt}
\renewcommand{\arraystretch}{1.5}
\begin{table}[htp]
\centering
\caption{Axial vector constant and flavor decomposition of $\Omega^{* 0}_{c}$} 
\label{tab:compareOmegacs}
\begin{threeparttable}
\resizebox{\columnwidth}{!}
{
 \begin{tabular}{ c | c c c c c c c c} 
\hline 
 \hline
 $J_{3}'=3/2$ & $g^{(0)}_{A}$  &  $g^{(3)}_{A}$ & $g^{(8)}_{A}$ & $\Delta{u}$ & $\Delta{d}$ & $\Delta{s}$ & $\Delta{c}$ \\
  \hline 
$\Omega^{* 0}_{c}$&$ 2.245 $ & $0.000$  &  $-1.797$ 
& $-0.104$ & $-0.104$  &  $1.452$  & $1.000$ \\  
$\Omega^{* 0}_{c}$~\cite{Alexandrou:2016xok}$^{\dagger}$&$ 2.1961 \pm 0.0261 $ & $-$  &  $-2.5817 \pm 0.0408^{o}$ 
& $-$ & $-$  &  $1.2904 \pm 0.0204$  & $0.9026 \pm 0.0090$ \\
    \hline 
 \hline
 \end{tabular} 
}
\begin{tablenotes}\footnotesize
\item[o] Note that the expressions for the axial charges in 
Ref.~\cite{Alexandrou:2016xok} are different from the present one 
by $\sqrt{3}$.
\item[$\dagger$] The values in Ref.~\cite{Alexandrou:2016xok} are
  obtained by averaging over various isospin partners.
\end{tablenotes}
\end{threeparttable}   
\end{table}

\end{document}